\documentclass[aps,prd,twocolumn,amsmath,showpacs,superscriptaddress,nofootinbib,nopreprintnumbers,tightenlines]{revtex4-1}

\usepackage{verbatim}
\usepackage[T1]{fontenc}
\usepackage[utf8]{inputenc}
\usepackage[american]{babel}
\usepackage{epsfig}
\usepackage{graphicx}

\usepackage{hyperref}

\usepackage{booktabs}
\usepackage{multirow}
\usepackage{dcolumn}
\usepackage{amsmath}
\usepackage{mathtools}
\usepackage{amsfonts}
\usepackage{amssymb}
\usepackage{epstopdf}
\usepackage{bm}
\usepackage{siunitx}
\usepackage{braket}
\usepackage{enumitem}
\usepackage{soul}
\usepackage[table]{xcolor}
\usepackage{color}
\usepackage{transparent}
\usepackage{pifont}

\definecolor{navyblue}{rgb}{0.0, 0.0, 0.5}
\definecolor{royalblue}{rgb}{0.25, 0.41, 0.88}
\definecolor{cadmiumgreen}{rgb}{0.0, 0.42, 0.24}
\definecolor{blue-violet}{rgb}{0.54, 0.17, 0.89}
\definecolor{darkviolet}{rgb}{0.58, 0.0, 0.83}
\definecolor{teal(colorwheel)}{rgb}{1.0, 0.5, 0.0}

\usepackage{hyperref}
\hypersetup{
    colorlinks=true, 
    linkcolor=royalblue, 
    citecolor=magenta}

\newcommand\ee{\end{equation}}
\newcommand\be{\begin{equation}}
\newcommand\eea{\end{eqnarray}}
\newcommand\bea{\begin{eqnarray}}

\renewcommand\({\left(}

\renewcommand\[{\left[}

\usepackage{booktabs}
\usepackage{multirow}
\usepackage{dcolumn}
\usepackage{colortbl}

\definecolor{magenta(process)}{rgb}{1.0, 0.0, 0.56}

\definecolor{darkspringgreen}{rgb}{0.09, 0.45, 0.27}

\definecolor{royalblue(web)}{rgb}{0.25, 0.41, 0.88}
\newcommand{\nq}[1]{%
	\begin{tabular}{@{}c@{}}\strut#1\strut\end{tabular}%
}

\begin{document}

\title{Cosmological Bound on the QCD Axion Mass, Redux}


\author{Francesco D'Eramo}
\email{francesco.deramo@pd.infn.it}
\affiliation{Dipartimento di Fisica e Astronomia, Universit\`a degli Studi di Padova, Via Marzolo 8, 35131 Padova, Italy}
\affiliation{Istituto Nazionale di Fisica Nucleare (INFN), Sezione di Padova, Via Marzolo 8, 35131 Padova, Italy}

\author{Eleonora Di Valentino}
\email{e.divalentino@sheffield.ac.uk}
\affiliation{School of Mathematics and Statistics, University of Sheffield, Hounsfield Road, Sheffield S3 7RH, United Kingdom}

\author{William Giar\`e}
\email{william.giare@uniroma1.it}
\affiliation{Galileo Galileo Institute for theoretical physics, Centro Nazionale INFN di Studi Avanzati, \\ Largo Enrico Fermi 2,  I-50125, Firenze, Italy}
\affiliation{Istituto Nazionale di Fisica Nucleare (INFN), Sezione di Roma, P.le A. Moro 2, I-00185, Roma, Italy}

\author{Fazlollah Hajkarim}
\email{fazlollah.hajkarim@pd.infn.it}
\affiliation{Dipartimento di Fisica e Astronomia, Universit\`a degli Studi di Padova, Via Marzolo 8, 35131 Padova, Italy}
\affiliation{Istituto Nazionale di Fisica Nucleare (INFN), Sezione di Padova, Via Marzolo 8, 35131 Padova, Italy}

\author{Alessandro Melchiorri}
\email{alessandro.melchiorri@roma1.infn.it}
\affiliation{Physics Department, Universit\`a di Roma ``La Sapienza'', Ple Aldo Moro 2, 00185, Rome, Italy}
\affiliation{Istituto Nazionale di Fisica Nucleare (INFN), Sezione di Roma, P.le A. Moro 2, I-00185, Roma, Italy}

\author{Olga Mena}
\email{omena@ific.uv.es}
\affiliation{IFIC, Universidad de Valencia-CSIC, 46071, Valencia, Spain}

\author{Fabrizio Renzi}
\email{renzi@lorentz.leidenuniv.nl}
\affiliation{Lorentz Institute for Theoretical Physics, Leiden University, PO Box 9506, Leiden 2300 RA, The Netherlands}

\author{Seokhoon Yun}
\email{seokhoon.yun@pd.infn.it}
\affiliation{Dipartimento di Fisica e Astronomia, Universit\`a degli Studi di Padova, Via Marzolo 8, 35131 Padova, Italy}
\affiliation{Istituto Nazionale di Fisica Nucleare (INFN), Sezione di Padova, Via Marzolo 8, 35131 Padova, Italy}

\date{\today}


\begin{abstract}
We revisit the joint constraints in the mixed hot dark matter scenario in which both thermally produced QCD axions and relic neutrinos are present. Upon recomputing the cosmological axion abundance via recent advances in the literature, we improve the state-of-the-art analyses and provide updated bounds on axion and neutrino masses. By avoiding approximate methods, such as the instantaneous decoupling approximation, and limitations due to the limited validity of the perturbative approach in QCD that forced to artificially divide the constraints from the axion-pion and the axion-gluon production channels, we find robust and self-consistent limits. We investigate the two most popular axion frameworks: KSVZ and DFSZ.  
From Big Bang Nucleosynthesis (BBN) light element abundances data we find for the KSVZ axion $\Delta N_{\rm eff}<0.31$ and an axion mass bound $m_a < 0.53 $~eV (i.e., a bound on the axion decay constant $f_a > 1.07 \times 10^7$~GeV) both at $95\%$~CL. These BBN bounds are improved to $\Delta N_{\rm eff}<0.14$ and $m_a< 0.16$~eV ($f_a > 3.56 \times 10^7$~GeV) if a prior on the baryon energy density from Cosmic Microwave Background (CMB) data is assumed. When instead considering cosmological observations from the CMB temperature, polarization and lensing from the Planck satellite combined with large scale structure data we find $\Delta N_{\rm eff}<0.23$, $m_a< 0.28$~eV ($f_a > 2.02 \times 10^7$~GeV) and $\sum m_\nu < 0.16$~eV at $95\%$~CL. This corresponds approximately to a factor of $5$ improvement in the axion mass bound with respect to the existing limits. Very similar results are obtained for the DFSZ axion. We also forecast upcoming observations from future CMB and galaxy surveys, showing that they could reach percent level errors for $m_a\sim 1$~eV.
\end{abstract}

\maketitle

\section{Introduction}
\label{sec.Intro}

Light and elusive degrees of freedom are ubiquitous in theoretical frameworks aimed at solving deficiencies of the standard model (SM) of particle physics. The feeble couplings and tiny masses imply a potentially prominent role in the early Universe, and their presence at early times can leave an imprint on cosmological observables. Some particle candidates for physics beyond the SM, such as the one investigated in this work, are somewhat special since they solve more than one problem.

Attempts to detect CP violation by strong interactions have not been successful so far. From the theory point of view, the source of CP violation (or time-reversal violation in agreement with the CPT theorem) in Quantum ChromoDynamics (QCD) is controlled by a dimensionless parameter $\theta$ that is expected to be of order one. The non-observation of CP-violating electric dipole moments of nucleons puts the spectacular bound $\theta \lesssim 10^{-10}$~\cite{Baker:2006ts,Pendlebury:2015lrz,Abel:2020pzs}, and understanding this severe inequality is at the heart of the so-called strong CP problem. 

An elegant solution is the Peccei-Quinn (PQ) mechanism~\cite{Peccei:1977hh,Peccei:1977ur} where the $\theta$ parameter is promoted to a dynamical pseudo-scalar field known as the \textit{axion}~\cite{Wilczek:1977pj,Weinberg:1977ma}. QCD dynamics itself drives such a field toward the configuration of minimal energy that is CP conserving~\cite{Vafa:1984xg}. Remarkably, the QCD axion kills two birds with one stone since it provides a solution to another serious SM drawback: the energy density stored in the field oscillations has a redshift behavior with the Hubble expansion which is the same as the one for non-relativistic matter~\cite{Preskill:1982cy,Abbott:1982af,Dine:1982ah}. Thus the QCD axion is one of the strongest motivated viable dark matter candidates, and the production in the early Universe is of non-thermal origin~\cite{Kibble:1976sj,Vilenkin:1981kz,Kibble:1982dd,Sikivie:1982qv,Vilenkin:1982ks,Linde:1985yf,Huang:1985tt,Seckel:1985tj,Davis:1986xc,Lyth:1989pb,Linde:1990yj,Vilenkin:2000jqa}.
 
Axions can be copiously produced in the early Universe also via scattering and decays of particles belonging to the primordial bath~\cite{Turner:1986tb,Berezhiani:1992rk,Brust:2013ova,Baumann:2016wac,DEramo:2018vss,Arias-Aragon:2020qtn,Arias-Aragon:2020shv,Green:2021hjh,DEramo:2021usm}. Eventually, once the early Universe gets cold enough and diluted, they propagate without any interaction along geodesics and free-stream until the present time. Their presence may alter the Hubble expansion rate since they contribute to the radiation energy density of the Universe as massive neutrinos do when they are still relativistic. Eventually, they become non-relativistic and provide a hot and sub-dominant dark matter component. The presence of such a cosmic axion background can leave distinct and detectable imprints, and current cosmological data put bounds on the axion mass and interactions~\cite{Hannestad:2005df,Melchiorri:2007cd,Hannestad:2007dd,Hannestad:2008js,Hannestad:2010yi,Archidiacono:2013cha,Giusarma:2014zza,DiValentino:2015zta,DiValentino:2015wba,Archidiacono:2015mda}. Recent work~\cite{Dror:2021nyr} suggested strategies to detect this population of axions even directly with terrestrial experiments but only for non-thermal production mechanisms. Axion couplings are proportional to the mass of the axion itself, and thermal production channels are efficient only if the axion mass is large enough. On the contrary, the cold axion dark matter density is a decreasing function of the mass. As a result, we have a significant thermal  population only if cold axions provide a sub-dominant component to the observed cosmic dark matter abundance. 

In this work, we revisit the joint constraints on the QCD thermal axion and neutrino masses and improve the state-of-the-art analysis~\cite{Giare:2020vzo}. First, we do not rely upon the instantaneous decoupling approximation. This method assumes that thermal equilibrium is achieved in the early Universe but this is not always true. Furthermore, the effective number of bath degrees of freedom changes substantially below the weak scale, and the resulting axion amount would be quite sensitive to the precise value of the decoupling temperature. We overcome these limitation with the Boltzmann equation formalism that tracks the axion population across the expansion history. The axion coupling to gluons, which is necessary to solve the strong CP problem, constitues a serious challenge since such a perturbative description breaks down once strong interactions confine. Historically, the axion production rate was evaluated exclusively above~\cite{Masso:2002np,Graf:2010tv,Salvio:2013iaa,Ferreira:2018vjj} or below~\cite{Chang:1993gm,DEramo:2014urw,Ferreira:2020bpb,DiLuzio:2021vjd,Carenza:2021ebx} the confinement scale, and cosmological bounds were derived by considering axion-pion and axion-gluon scatterings separately~\cite{Giare:2020vzo}. Refs.~\cite{DEramo:2021psx,DEramo:2021lgb} provided continuous results for the axion production rate. In the high-temperature regime where perturbative QCD is trustworthy, a proper thermal field theory treatment regulates the singular IR behavior due to the long-range interactions mediated by gluons. At low temperatures, axions are produced via pion collisions. The two results are connected via a smooth interpolation across the QCD crossover. We employ these rates in our analysis, and the limits obtained here are more constraining than those previously found in the literature.\footnote{The recent analysis in Ref.~\cite{Caloni:2022uya} employed the rates provided by Refs.~\cite{DEramo:2021psx,DEramo:2021lgb} to bound thermally-produced axion-like particles.}

Section~\ref{sec.axion} describes the QCD axion scenarios explored here, whose thermal production mechanism is detailed in Sec.~\ref{sec.production}. The cosmological analyses start in Sec.~\ref{sec.Cosmo}, which include a detailed description of the numerical implementation. Section~\ref{sec.Results} presents our results, derived using Big Bang Nucleosynthesis (BBN) and current cosmological measurements. Forecasted analyses based in future Cosmic Microwave Background (CMB) and galaxy survey observations will also be shown. We conclude in Sec.~\ref{sec:conclusions}.

\section{QCD axion: mass and interactions}
\label{sec.axion}

Every QCD axion model features a global Abelian $U(1)$ symmetry, also known as PQ symmetry, which is anomalous under strong interactions and spontaneously broken. The QCD axion is the pseudo-Nambu-Goldstone-Boson (PNGB) that originates from such a  breaking. The color anomaly of the PQ symmetry is responsible for a tiny axion mass that has profound cosmological consequences. 

Regardless of the specific microscopic realization, the color anomaly manifests itself at low energy through the contact interactions of the axion field $a$ with gluons
\be
\mathcal{L}_{aG} = \frac{\alpha_s}{8 \pi} \frac{a}{f_a} G^A_{\mu\nu} \widetilde{G}^{A \mu\nu} \ .
\label{eq:LPQaxion}
\ee
We define $\alpha_s = g_s^2 / (4 \pi)$ the QCD fine structure constant, $G_{\mu\nu}^A$ the gluon field strength, $\widetilde{G}^{A \mu\nu} \equiv \epsilon^{\mu\nu\rho\sigma} G^A_{\rho\sigma} / 2$ its dual, and $f_a$ is known as the axion decay constant. Once we reach the confinement scale, strong QCD dynamics generate a non-perturbative potential that leads to the general expression for the axion mass~\cite{Bardeen:1978nq,GrillidiCortona:2015jxo,Gorghetto:2018ocs}
\be
m_a \simeq 5.7 \, \mu{\rm eV} \, \left( \frac{10^{12} \, {\rm GeV}}{f_a} \right) \ .
\label{eq:ma}
\ee

Other axion couplings are allowed but not mandatory. Anomalous interactions to electroweak gauge bosons take a form similar to the one for gluons with different gauge couplings and model-dependent Wilson coefficients. The axion can couple via derivative interactions to SM fermions that we write schematically as follows
\be
\mathcal{L}_{a \psi} = \frac{\partial_\mu a}{f_a} \,
\overline{\psi^\prime} \gamma^\mu \left(c^V_{\psi^\prime \psi} + c^A_{\psi^\prime \psi} \gamma^5 \right) \psi \ .
\label{eq:apsicoupling}
\ee
The fields $\psi$ and $\psi^\prime$ do not necessarily belong to the same generation, and electroweak gauge invariance relates vector and axial-vector couplings. The derivative coupling preserves the shift symmetry $a \rightarrow a + {\rm const}$ consistently with the PNGB nature of the axion. 

The landscape of axion models is broad~\cite{Kim:2008hd,DiLuzio:2020wdo}. We divide them into two main classes according to the origin of the color anomaly: Kim-Shifman-Vainshtein-Zakharov (KSVZ)~\cite{Kim:1979if,Shifman:1979if} and Dine-Fischler-Srednicki-Zhitnitsky (DFSZ)~\cite{Dine:1981rt,Zhitnitsky:1980tq} frameworks. The coupling to gluons in Eq.~\eqref{eq:LPQaxion} and the consequent axion mass in Eq.~\eqref{eq:ma} are the same for both cases. Interactions with electroweak gauge bosons and SM fermions in Eq.~\eqref{eq:apsicoupling} are different instead.

\subsection{KSVZ Axion}
\label{subsec.KSVZ}

None of the SM fields carries a PQ charge in this case. New colored fermions are responsible for the axion interaction with gluons. They get their mass from PQ breaking, and they have to be heavier than the TeV range to satisfy the collider bounds on new colored particles. For processes at energies below this mass, the new fermions can be integrated out and this procedure generates the local operator in Eq.~\eqref{eq:LPQaxion}. Anomalous couplings to electroweak gauge bosons can be present depending on the gauge charges of these fermions, but their effect is sub-dominant given the hierarchy among gauge coupling constants. Interactions with SM fermions are absent. 

We are interested in axion production at temperatures below the weak scale\footnote{Ref.~\cite{DEramo:2021lgb} provides the axion production rate across the intermediate mass threshold of the heavy PQ charged fermions.} so we employ the low-energy description in Eq.~\eqref{eq:LPQaxion}. Gluon scatterings are responsible for production above the GeV scale until the epoch when quarks and gluons get confined into hadrons. Below confinement, axion production is controlled by pion scatterings via the derivative interaction
\be
\mathcal{L}_{a\pi\pi\pi} = \frac{c_{a\pi\pi\pi}}{f_\pi} \, \frac{\partial_\mu a}{f_a} \,  \mathcal{J}_\pi^\mu \ .
\label{eq:pionKSVZ}
\ee
Here, $f_\pi \simeq 93 \, {\rm MeV}$ is the pion decay constant and we define the spin-one pion current
\be
\mathcal{J}_{\pi}^\mu \equiv \pi^0\pi^+\partial^\mu \pi^- + \pi^0\pi^-\partial^\mu \pi^+ - 2 \pi^+\pi^-\partial^\mu \pi^0 \ .
\ee
The only model dependent coupling is the dimensionless Wilson coefficient $c_{a\pi\pi\pi}$ that in our case results in 
\be
c_{a\pi\pi\pi}^{\rm KSVZ} \simeq 0.12 \ .
\ee
 
\subsection{DFSZ Axion}
\label{subsec.DFSZ}

There is no new fermion field in this case, and the color anomaly of the PQ symmetry is generated by SM quarks that carry a non-vanishing PQ charge. The Higgs sector needs to be extended with another weak doublet and this brings an additional intermediate scale in the game: the mass of the heavy Higgs bosons. We work in the so-called decoupling limit, which is motivated also by collider bounds, where such a mass is well-above the weak scale. In particular, the heavy Higgs bosons mass is above the energies that we consider for axion production.\footnote{Ref.~\cite{DEramo:2021lgb} provides the axion production rate across the intermediate mass threshold of the heavy Higgs bosons.}

The local operator in Eq.~\eqref{eq:LPQaxion} arises once we account for the anomalous effects of all SM quarks. Thus it is non-local above the weak scale since the top quark is the heaviest PQ-charged and colored fermion. At intermediate scales, the contact interaction is present with a scale-dependent anomaly coefficient accounting for the virtual effects of heavy quarks that have been integrated out.

The presence of two Higgs doublets $H_u$ and $H_d$ introduces a new parameter relevant to axion phenomenology: the ratio between their vacuum expectations values (vevs) $v_u$ and $v_d$. We follow the standard convention and parameterize this ratio as $\tan\beta = v_u/ v_d$.  These vevs generate fermion masses, and we consider a type-II two Higgs doublet model (2HDM). Unlike the previous case, the axion couples to SM fermions since they carry a PQ charge. The parameter $\tan\beta$ sets the relative interaction strength between different fermions. Thermal axion production is possibly sensitive to the value of $\tan\beta$ since fermion scattering is an active production source. 

Axion production in this scenario is richer. Above confinement, we have scattering processes caused by the interactions with SM fermions in Eq.~\eqref{eq:apsicoupling} besides quark and gluon scatterings mediated by the operator in Eq.~\eqref{eq:LPQaxion}. Below the confinement scale, axion production via electron and muon scatterings is present in this case. Concerning pion scatterings, axion production is controlled again by the interactions in Eq.~\eqref{eq:pionKSVZ}. However, the additional axion couplings to SM quarks modify the effective dimensionless parameter $c_{a\pi\pi\pi}$ as follows 
\be
c_{a\pi\pi\pi}^{\rm DFSZ} = c_{a\pi\pi\pi}^{\rm KSVZ} - \frac{\cos 2\beta}{9} \ .
\ee

\section{Thermal Production}
\label{sec.production}

Scatterings are the only production channel available for the minimal KSVZ and DFSZ frameworks due to the flavor-conserving axion interactions. If flavor-violating couplings are allowed then decays also contribute~\cite{DEramo:2021usm}. Initially, axions may belong to the thermal bath if during or after inflationary reheating there are efficient mechanisms to produce them. Even if they are not present at the onset of the radiation domination epoch, they are produced by thermal scatterings and their interaction strength could bring them to equilibrium. Whether they thermalize or not, there will be a moment starting from which interactions are so rare that axions just free-stream until the present time. Keeping track of this axion population through the expansion history is the goal of this section. The output of this procedure is the prediction of the number of axions in a comoving volume as a function of the axion mass, and this is also the input for the cosmological analysis performed in the remaining part of this work.

A quick method to estimate the axion relic density is via the \textit{instantaneous decoupling approximation}. Here, we assume that axions at some point reach thermal equilibrium, and that they suddenly decouple when the bath had a temperature $T_D$ identified by the relation 
\be
\Gamma_a(T_D) = H(T_D) \ .
\label{eq:TD}
\ee
The temperature dependent production rate $\Gamma_a(T)$ depends on the axion model and it quantifies the number of production processes that happen per unit time. On the other side of the equality, we have the Hubble expansion rate set by the Friedmann equation
\be
H(T) = \frac{\sqrt{\rho_{\rm rad}}}{\sqrt{3} M_{\rm Pl}} = 
\frac{\pi \sqrt{g_\star(T)}}{3 \sqrt{10}} \frac{T^2}{M_{\rm Pl}} \
\label{eq:Friedmann}
\ee
with $M_{\rm Pl} \equiv (8 \pi G)^{-1/2}  \simeq 2.44\times10^{18}\, {\rm GeV}$ the reduced Planck's mass. The last equality holds for a radiation dominated Universe with a temperature dependent $g_\star(T)$ relativistic effective degrees of freedom. Axions just free-stream for temperatures below $T_D$ and therefore their comoving number density $Y_a \equiv n_a / s$ stays constant. Here, $s$ is the entropy density of the thermal bath which is conventionally parameterized as $s=(2\pi^2/45)g_{\star s}(T)T^3$ with $g_{\star s}(T)$ the number of entropic degrees of freedom. The asymptotic axion comoving density results in
\be
Y_a^\infty = Y_a(T \lesssim T_D) = \frac{45 \zeta(3)}{2 \pi^4 g_{\star s}(T_D)} 
\label{eq:Yainf}
\ee
with $\zeta(3) \simeq 1.2$ the Riemann zeta function. The asymptotic  comoving density depends on the decoupling temperature only through the factor $g_{\star s}(T_D)$ in the denominator. Axions decoupling at later times are a more significant fraction of the thermal bath and therefore their comoving density is larger. 

\begin{figure}
	\centering
	\includegraphics[width=\columnwidth]{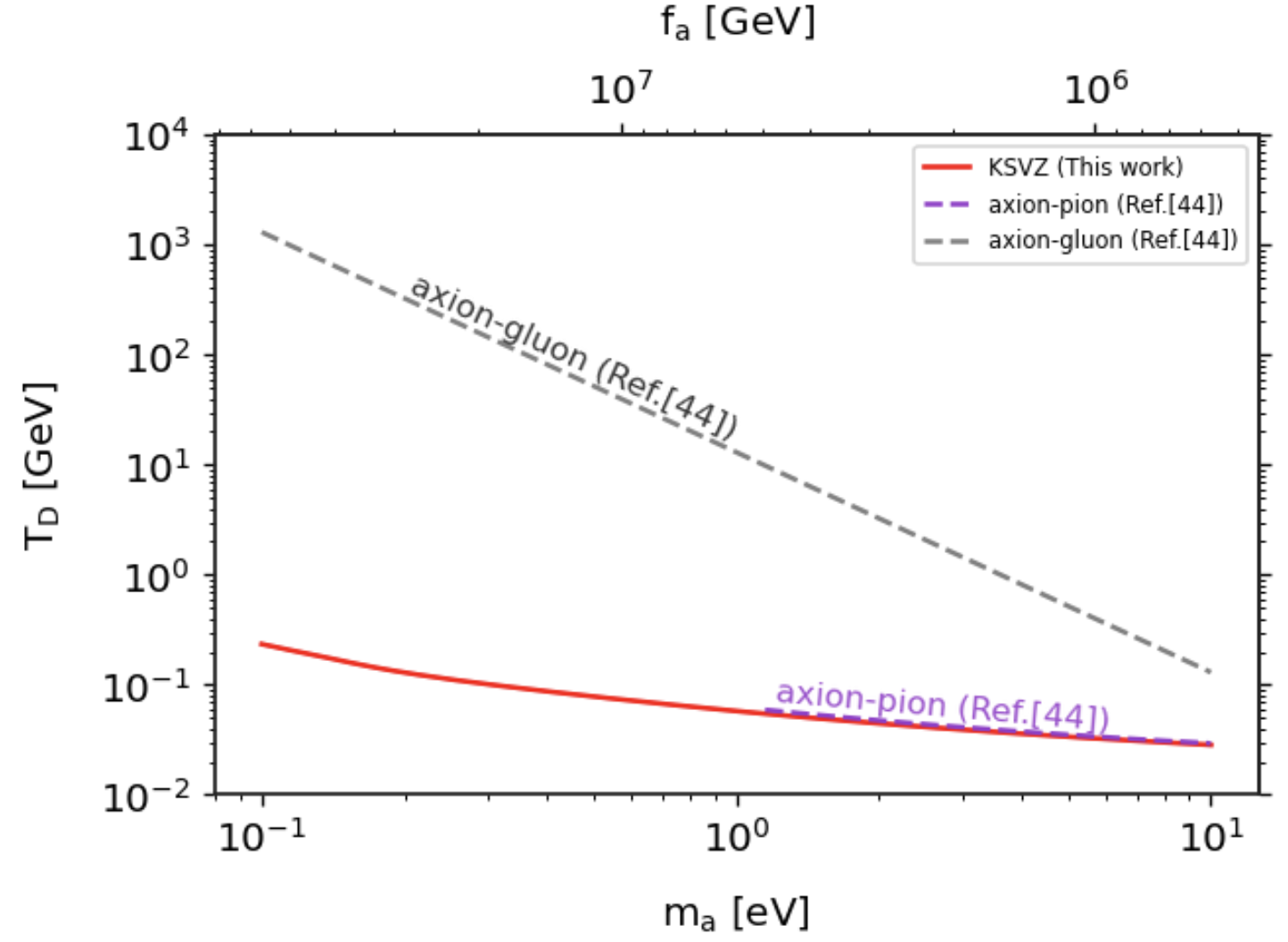}
	\caption{Decoupling temperature $T_D$ as a function of the axion mass $m_a$ (lower horizontal axis) and the axion decay constant $f_a$ (upper horizontal axis). Dashed lines corresponds to the values found by Ref.~\cite{Giare:2020vzo}, the solid line corresponds to the KSVZ axion with a continuous result for the rate.}
	\label{fig:TD}
\end{figure}

\autoref{fig:TD} shows the decoupling temperature obtained via Eq.~\eqref{eq:TD} as a function of the  axion mass (or, equivalently, as a function of the axion decay constant $f_a$ through the relation in Eq.~\eqref{eq:ma}) for the KSVZ axion via the production rates of Refs.~\cite{DEramo:2021psx,DEramo:2021lgb}. We use the $g_{\star}(T)$ and $g_{\star s}(T)$ functions provided by Ref.~\cite{Saikawa:2018rcs}, and we check that the outcome of our cosmological analysis is not sensitive to this specific choice.\footnote{We checked that the mass bounds obtained is this work are the same that one would obtain with the functions in Refs.~\cite{Drees:2015exa,Laine:2015kra}} For comparison, we reproduce also the results of Ref.~\cite{Giare:2020vzo} that estimated the axion relic abundance for the KSVZ axion within the instantaneous decoupling approximation by separating two different regimes: gluon and pion scatterings. This corresponds to the dashed lines in the figure. For axion-gluon scattering, the decoupling temperature was significantly overestimated.

The pion rate employed by Ref.~\cite{Giare:2020vzo} results from a standard calculation within chiral perturbation theory and therefore is reliable. This is why the axion-pion dashed line agrees well with the full KSVZ estimate of the decoupling temperature. On the contrary, the axion-gluon dashed line is significantly far away from the correct KSVZ result. The production rate adopted by Ref.~\cite{Giare:2020vzo} in the deconfined phase corresponds to a dimensional analysis estimate. A proper rate evaluation that accounts for the running of the strong coupling constant enhances the rate as we approach the QCD crossover, and therefore it reduces the decoupling temperature. The net result is an underestimate of the axion relic density by previous studies, and this is the reason why the cosmological mass bounds found in this analysis are stronger. Therefore, we improve in this manuscript the results presented in  Ref.~\cite{Giare:2020vzo}, which regardless if there were derived relying in estimated rates, have pioneered cosmological axion constraints and motivated studies as the one presented here.

The criterion in Eq.~\eqref{eq:TD} leads to a ballpark estimate of the decoupling epoch but it is far from being rigorous. As it is manifest from Eq.~\eqref{eq:Yainf}, the final axion abundance is sensitive to the detailed value of the decoupling temperature only if $g_{\star s}$ is changing around the decoupling time. This is exactly the case for the axion parameter space investigated in this work. Furthermore, the interaction rate can have a peculiar structure around the decoupling temperature as in DFSZ theories where threshold effects around the mass of SM heavy quarks enhances the axion coupling to gluons.\footnote{This is analogous to integrating out a heavy PQ-charged and colored fermion in KSVZ theories as discussed in detail by Ref.~\cite{DEramo:2021lgb}.}

We need to refine the axion relic density calculation in order to find solid cosmological bounds on the axion mass. A proper tool to follow the axion number density $n_a$ in the early Universe is the Boltzmann equation
\be
\frac{dn_{a}}{dt}+3Hn_{a}= \gamma_{a}\left(1-\frac{n_{a}}{n_{a}^{\rm eq}}\right)\ .
\ee
Here, $n_a^{\rm eq}$ is the axion equilibrium number density and $\gamma_{a}$ accounts for axion number-changing processes. The latter quantity is connected to the interaction rate $\Gamma_a$ introduced before via the relation $\gamma_a = n_a^{\rm eq} \Gamma_a$. We find it convenient to rewrite the Boltzmann equation in terms of dimensionless quantities. We switch to the comoving density $Y_a$ already introduced before, and we trade the time variable $t$ with the dimensionless combination $x\equiv M / T$. The overall mass scale $M$ is purely conventional and it can be chosen to make the numerical solution easier. The Boltzmann equation reads
\be
\frac{dY_{a}}{d\log x}=-\left(1-\frac{1}{3}\frac{d\log g_{\star s}}{d \log x}\right)\frac{\gamma_{a}(x)}{H(x)s(x)}\left(1-\frac{Y_{a}}{Y_{a}^{\rm eq}}\right) \ .
\ee
We solve this equation from an initial temperature at the TeV scale and we choose two boundary conditions: axions initially in thermal equilibrium, and vanishing initial axion population. As shown explicitly by Refs.~\cite{DEramo:2021psx,DEramo:2021lgb}, for the axion mass range studied here ($m_a \gtrsim 0.1 \, {\rm eV}$) the initial condition does not matter if we start the evolution above the weak scale. We solve the Bolzmann equation numerically down to temperatures well below the masses of particles participating in the production processes, and we find that the axion yield reaches a constant value $Y_a^{\infty}$.

\begin{figure}
	\centering
	\includegraphics[width=\columnwidth]{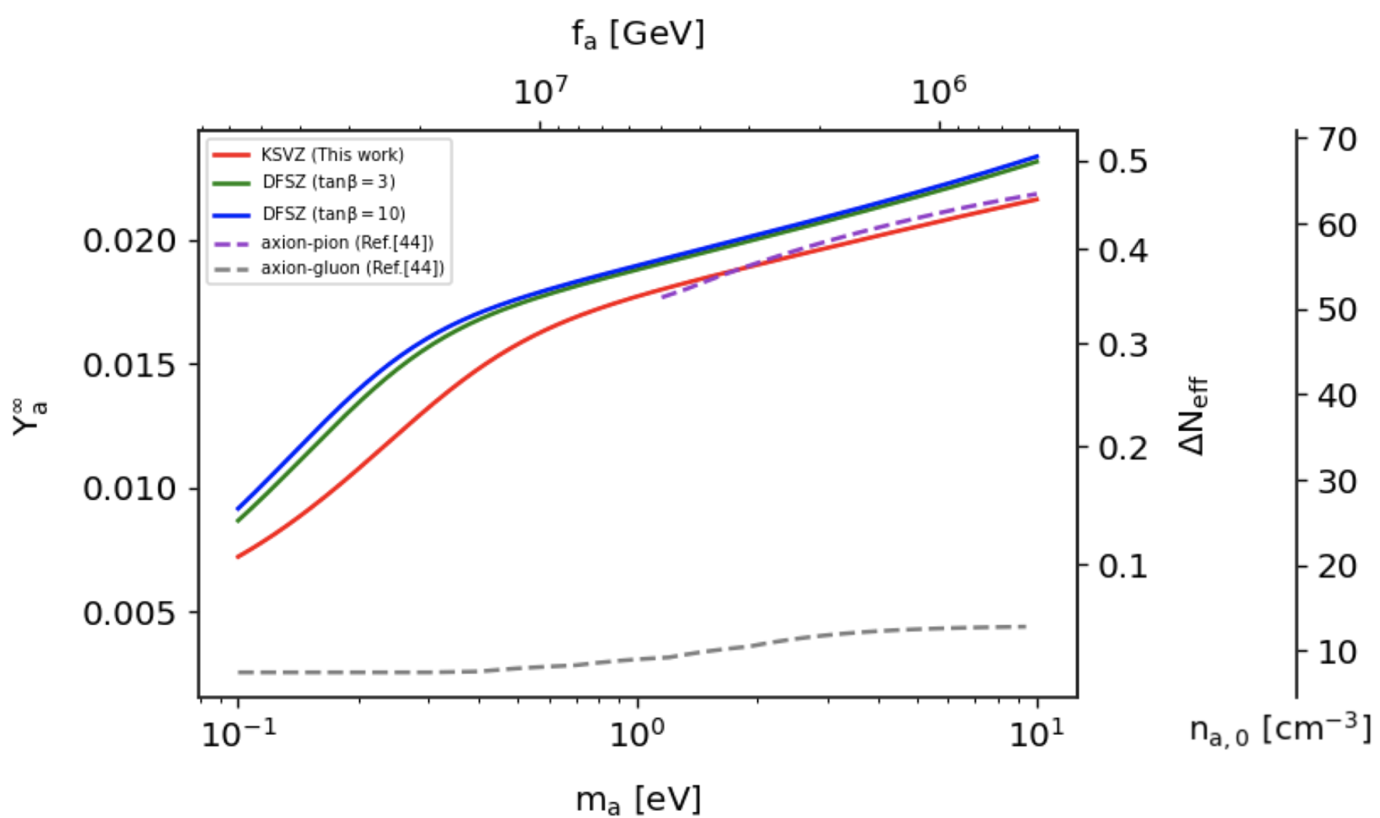}
	\caption{Asymptotic axion yield $Y_a^\infty$ as a function of $m_a$ (lower horizontal axis) and $f_a$ (upper horizontal axis). The right vertical axis identifies the correspondent values of $\Delta N_{\rm eff}$ and $n_{a,0}\equiv n_a(T_0)$. Solid lines are obtained via a Boltzmann equation solution. We show with dashed lines also the results of Ref.~\cite{Giare:2020vzo} for axion couplings to gluons and pions.}
	\label{fig:maYa}
\end{figure}

\autoref{fig:maYa} shows the final axion comoving density for both the KSVZ and the DFSZ cases, and for the latter we choose two different values of $\tan\beta$. For comparison, we show again the result of the previous analysis in Ref.~\cite{Giare:2020vzo} for the couplings to gluons and pions. The actual axion relic density is substantially larger than previously estimated. We also notice how the value of $\tan\beta$ does not alter the resulting axion abundance significantly, and in what follows we fix $\tan\beta = 3$ for the DFSZ axion case.

Finally, we evaluate the axion contribution to the effective number of additional neutrino species via the relation $\Delta N_{\rm eff}\simeq 75.6\,\left(Y_a^{\infty}\right)^{4/3}$; this results from the axion number density conversion into the correspondent energy density under the assumption that the phase-space distribution is thermal. Such a procedure is justified by the analysis in Ref.~\cite{DEramo:2021lgb} where the authors show how axions with mass in the range considered in this analysis always achieve thermalization and then decouple eventually. Crucially, our Boltzmann equation methodology fully captures the dynamics at the decoupling epoch and we are able to track the axion population rigorously before, after, and during the freeze-out of interactions.

\section{Cosmological Analysis}
\label{sec.Cosmo}

As aforementioned in the introduction, QCD axions thermally produced in the early Universe can leave distinctive signatures in different cosmological observables that we shall review in what follows. We study the Universe at the times of BBN and recombination where the axions are not in thermal equilibrium with the thermal bath anymore. For this reason, we cannot use a single bath temperature $T$ as done in the previous section but we need to distinguish among the different temperatures for photons ($T_\gamma$), neutrinos ($T_\nu$), and axions ($T_a$).

\subsection{Thermal Axion Cosmology}
\label{sec.Cosmo.Theory}

As along as thermal axions remain relativistic particles ($T_a\gg m_a$), they behave as radiation in the early Universe and their cosmological effects are those produced via their contribution to the effective number of neutrino species $N_{\rm eff}$. We recall its definition via the energy density of radiation as a function of the redshift $z$
\begin{equation}
\rho_{\rm rad}(z)=\frac{\pi^2}{30}\left[ 2 T^4_{\gamma}(z) + \frac{7}{4} N_{\rm eff} T^4_{\nu}(z)\right]~ .
\end{equation}
Light relativistic axions will provide an additional contribution $\Delta \rho_{\rm rad}(z)\propto T_a^4(z)$ that we parameterize in terms of a correction $\Delta N_{\rm eff} \equiv N_{\rm eff}-3.044$ to the effective number of neutrino species with respect to the SM.\footnote{We recall that the effective number of neutrino species in the SM is $N^{\rm SM}_{\rm eff}=3.044$, see Ref.~\cite{Bennett:2020zkv} and references therein.} By means of the Boltzmann equation outlined in the previous section, it is possible to evaluate this correction and retrace the effects of relativistic axions through the different cosmological epochs. 

Current limits on $\Delta N_{\rm eff}$ arise primarily from observables at two epochs: \textit{(i)} at the BBN period, and, \textit{(ii)} at the Recombination epoch. 

BBN is a cornerstone of Hot Big Bang cosmology since it explains the formation of the first light nuclei (heavier than the lightest isotope of hydrogen) by a solid understanding of the nuclear interactions involved in the production of elements.  It also provides a natural laboratory to test extensions of the SM of elementary particles that involve additional relativistic species. Indeed, additional contributions to $\Delta N_{\rm eff}$ will increase the expansion rate $H(z)$ by Eq.\eqref{eq:Friedmann} and lead to a higher freeze-out temperature of the weak interactions, implying a higher fraction of primordial Helium and Deuterium (as well as a higher fraction of other primordial elements). This makes BBN a powerful tool for constraining the total amount of relativistic species and beyond the SM physics frameworks: given a concrete model, we can solve numerically the set of differential equations that regulate the nuclear interactions in the primordial plasma~\cite{PitrouEtal2018,Pisanti:2007hk,Consiglio:2017pot,Gariazzo:2021iiu}, compute the light element abundances and compare the results to the values inferred by astrophysical and cosmological observations. Given current uncertainties, the standard BBN predictions show a good agreement with direct measurements of primordial abundances~\cite{Pitrou:2020etk,Mossa:2020gjc,Pisanti:2020efz,Yeh:2020mgl} limiting $\Delta N_{\rm eff}\lesssim 0.3-0.4$ at 95\% CL. Notice also that BBN is expected to start at temperatures of the order of 1 MeV (after the neutrino decoupling) and end well before the recombination epoch. Therefore this bound can be safely applied to all thermal axions that are relativistic in this range of temperatures and it is highly independent of recombination processes. Conversely, the BBN predictions for the Helium abundance can impact the CMB angular spectra because they can be used to estimate the baryon energy density through a simple formula~\cite{Serpico:2004gx}
\be
    \Omega_b h^2 = \frac{1 - 0.007125\ Y_p^{\rm BBN}}{273.279}\left(\frac{T_{\rm CMB}}{2.7255\ \mathrm{K}}\right)^3 \eta_{10} \ .
    \label{eq.Yp}
\ee
Here, $\eta_{10} \equiv 10^{10}n_b/n_\gamma $ is the photon-baryon ratio today, $T_{\rm CMB} $ is the CMB temperature at the present time and $Y_p^{\rm BBN} \equiv 4 n_{\rm He}/n_{b}$ is the Helium \textit{nucleon fraction} defined as the ratio of the 4-Helium to the baryon density one.\footnote{We will provide results in term of the Helium \textit{mass fraction} defined as $Y_P = Y_{\rm{He4}} \,m_{\rm{He4}} /[Y_{\rm{He4}}\, m_{\rm{He4}} + (1 - 4\,Y_{\rm{He4}}) \,m_{\rm{H1}}]$ with $ Y_{\rm{He4}}=1/4 \,Y_{p}^{\rm {BBN}}$, $m_{\rm{He4}}\simeq 4.0026$, and $m_{\rm{H1}}\simeq 1.0078$.}

Considering lower temperatures, additional dark radiation at recombination leads to some characteristic signatures in the CMB temperature angular power spectrum, modifying  the damping tail and changing two important scales: the sound horizon and the Silk damping scale~\cite{Hou:2011ec,Archidiacono:2013fha}. The final release of Planck 2018 temperature and polarization data constrains $\Delta N_{\rm eff}\lesssim 0.4$ at 95\% CL ~\cite{Akrami:2018vks}, basically providing the same sensitivity on additional relativistic relics as BBN constraints. However, in this case, axions must behave as relativistic particles at recombination, i.e. $m_{a}\lesssim T_a(z_{\rm{LS}})$ (where $z_{\rm{LS}}\simeq 1100$ is the redshift corresponding to the last scattering surface). The temperature of a population of axion decoupled from the thermal bath at $T_{\rm D}$ at recombination is given by
\begin{equation}
T_a(z_{\rm{LS}}) \simeq T_{\rm rec} \times \left(\frac{2}{g_{\star}(T_{\rm D})}\right)^{1/3}~,
\label{eq:TaLS}
\end{equation}
with $T_{\rm rec}\simeq 0.3$ eV. Consequently only very light axions ($m_a\lesssim 0.1$~eV) are still relativistic during this epoch.
\\

When axions become non-relativistic they contribute to the total amount of (hot dark) matter of the Universe. We quantify this contribution via the parameter $\Omega_a$ that, in the non-relativistic regime, reads
\begin{equation}
	\Omega_a \, h^2 \simeq \frac{m_a\,n_a}{1.054\cdot 10^4\,\rm{eV}\,\rm{cm}^{-3}}~.
\end{equation}
Non-relativistic axions leave signatures similar to massive neutrinos, suppressing structure formation at scales smaller than their free-streaming length and leaving an imprint on the CMB temperature anisotropies via the early integrated Sachs-Wolfe effect. Therefore we expect the axion and total neutrino masses to be anti-correlated: increasing the axion mass leads to a larger hot dark matter component so that the contributions from massive neutrinos should be reduced to keep the total amount of hot dark matter consistent with the data. This is one of the reasons why multi-messenger analysis of the axion and neutrino effects on cosmological scales turned out to be a very promising tool for constraining their properties, also with direct implications for current and future direct experiments. Hence in the following analysis we study a realistic mixed hot dark matter scenario of axions and neutrinos taking also into account a possible non-vanishing neutrino mass, as robustly established by oscillation experiments~\citep{deSalas:2020pgw,deSalas:2018bym}. 

\subsection{Numerical Implementation}
\label{sec.Cosmo.Implementation}

\begin{figure}
	\centering
	\includegraphics[width=\columnwidth]{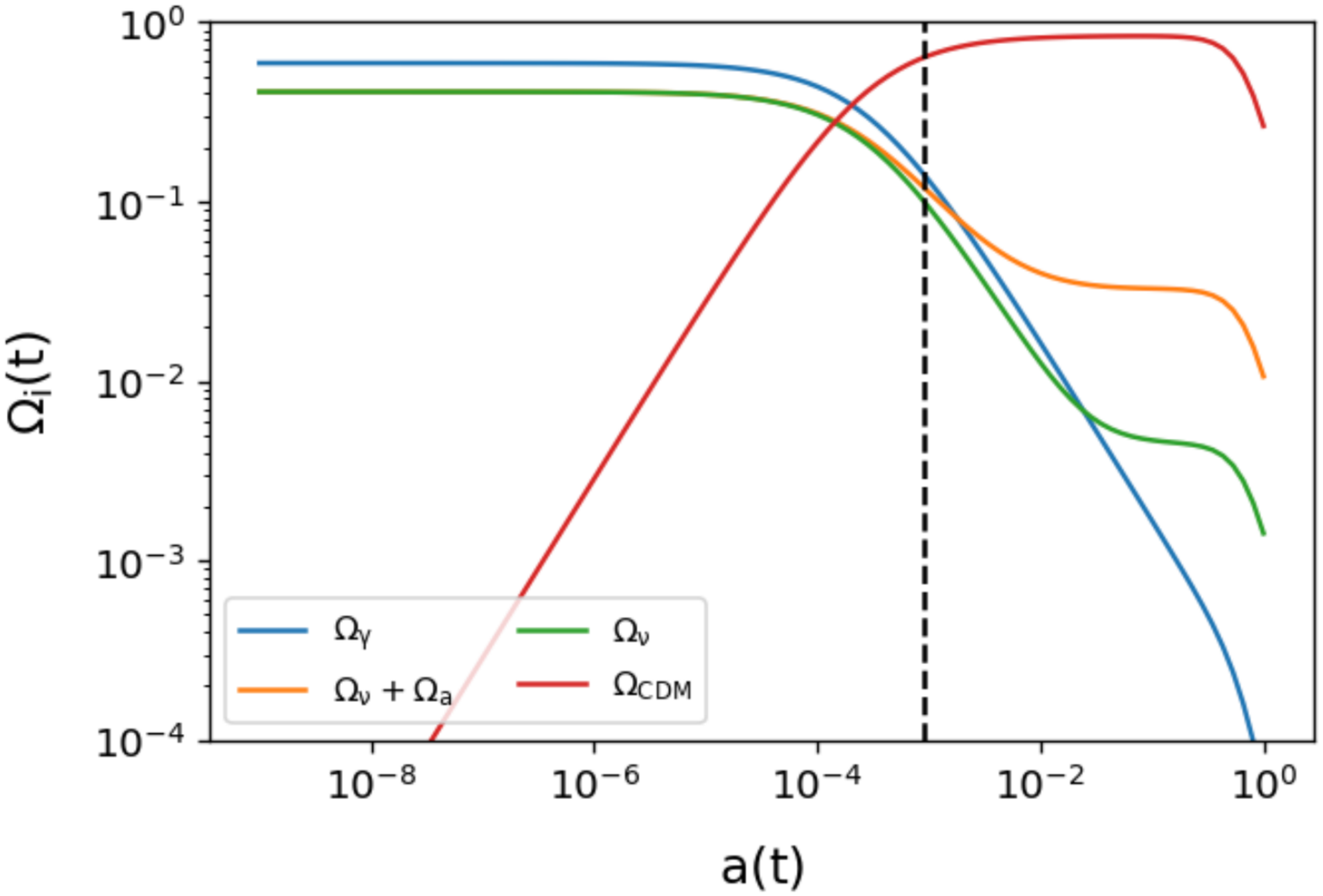}
	\caption{Energy density, normalized to the critical density, as a function of the scale factor for photons, massive neutrinos, axions and cold dark matter. An axion mass $m_a=1$~eV and a total neutrino mass $\sum m_\nu=0.06$~eV have been assumed (see the main text for details). The dashed vertical line represents approximately the value of the scale factor at recombination.}
	\label{fig:Density}
\end{figure}

In order to fully address the different effects induced by a relic population of thermal axions, we use a modified version of the latest version of the cosmological Boltzmann integrator code \textsc{CAMB} \citep{Lewis:1999bs,Howlett:2012mh}. In particular, the code has been modified to describe the QCD axion effects on cosmological scales only in terms of the axion mass that we include as an additional cosmological parameter in our analyses. For each value of the axion mass, the numerical calculation provides the axion contribution through various cosmological epochs, switching between the relativistic and non-relativistic regime (\autoref{fig:Density}). In what follows, we briefly retrace the modifications in the code and how we implemented them.  

The QCD axion mass window of interest for this analysis ranges from $0.1$ eV to $10$ eV.  Given current bounds on hot thermal relics (see e.g. Ref.~\cite{Giare:2020vzo}), the axion is always relativistic during the BBN, contributing as additional dark radiation and modifying the theoretical predictions for the abundance of primordial elements. We make use of the code \textsc{PArthENoPE}~\citep{Gariazzo:2021iiu} to quantify these effects. Starting from the input values of $N_\mathrm{eff}$, the neutron lifetime $\tau_n$ and the baryon energy density $\Omega_b h^2$, we evaluate the primordial light element abundances by solving numerically a set of differential equations that regulate the nuclear interactions in the primordial plasma after neutrino decoupling ($T \gtrsim 1$ MeV) until the end of BBN ($T \sim 10$ keV). We include the BBN predictions in our analyses, following the same procedure adopted by the Planck collaboration~\citep{Aghanim:2018eyx}: we fix the neutron lifetime  to $\tau_n = 879.4$ s, corresponding to the latest value reported by the Particle Data Group ($\tau_n = 879.4 \pm 0.6 $ s)~\citep{ParticleDataGroup:2020ssz} and create an interpolation grid varying $\Omega_b h^2 $ and $\Delta N_\mathrm{eff}$ within $\Delta N_\mathrm{eff} \in [0 ; 3]$ and $\Omega_b h^2 \in [0.0073 ; 0.033]$, respectively.  In this way, for each value of the axion mass, we compute the corresponding  $\Delta N_{\rm eff}(m_a)$ by solving the Boltzmann equation and the corresponding BBN predictions for all the light elements, from $\rm H$ up to $^7\rm Li$. 

Very light axions ($m_a\sim 0.1$ eV) are still relativistic at recombination and thus they modify the CMB angular spectra always trough their effects in $\Delta N_{\rm eff}$. Notice however that the correction to the effective number of relativistic particles coming from the tiny mass values aforementioned ($m_a\sim 0.1$ eV) is typically beyond the sensitivity of current CMB data, while it can be relevant for future CMB stage-4 experiments~\cite{Giare:2021cqr}.

Heavier axions, although relativistic during the BBN epoch, can be highly non-relativistic at recombination. In this case, they impact the CMB angular power spectra both indirectly (through their modification of the primordial Helium abundance during BBN) and directly (by the early integrated Sachs-Wolfe effect, increasing the amount of hot dark matter, similarly to massive neutrinos). We consider both effects. For the former one, a combined analysis of CMB and BBN data strongly disfavors contributions larger than $\Delta N_{\rm eff}\gtrsim 0.4$ (at both epochs) thus setting a robust upper limit on additional dark radiation in the early Universe. For the latter case, we dynamically evolve the axion energy density $\Omega_a(t)$ during all the cosmological evolution via the same procedure used for massive neutrinos and properly calculate the late-time effects on the matter power spectrum and structure formation. As an explicit example, \autoref{fig:Density} shows the energy density as a function of the scale factor for photons, axions, neutrinos and cold dark matter (assuming $m_a=1$~eV and $\sum m_\nu =0.06$~eV). Both neutrinos and axions behave as radiation at early times, and they become a sub-dominant dark matter contribution once they turn non-relativistic. For the specific mass values chosen to produce \autoref{fig:Density}, one can also appreciate that the contribution to the energy density of massive neutrinos and axions may differ significantly. In particular, axions with masses larger than $0.1$ eV become non-relativistic before the CMB time thus they start contributing to the matter density earlier (than neutrinos) leading to a significant impact on structure formation. As we shall see below, this feature will allow us to put stringent constraints on axion masses using large scale structure data. When the two species has similar masses, the evolution of their energy densities is preventing us to constrain axion masses that are smaller than the current bounds on neutrino masses $\sum m_\nu \lesssim 0.2 $ eV.

\section{Results}
\label{sec.Results}

\subsection{Results from BBN}
\label{sec.Results.BBN}
\begin{table*}
	\begin{center}
		\renewcommand{\arraystretch}{1.5}
		\resizebox{0.6\textwidth}{!}{\begin{tabular}{l  c  c  c}
		\multicolumn{4}{c}{}\\
		\multicolumn{4}{c}{\textbf{Constraints from BBN only}}\\[0.2cm]
  	        \hline
		    \textbf{\nq{Parameter \\}} & \textbf{\nq{BBN-A\\($Y_p + \Omega_b\,h^2$)}}  &  \textbf{\nq{BBN-B\\($Y_p + D/H$)}} &  \textbf{\nq{BBN-C\\($Y_p + D/H + \Omega_b\,h^2$)}} \\
		    \hline\hline
		\multicolumn{4}{c}{\textbf{KSVZ Axion Model}}\\
			\hline\hline
			$\Omega_{\rm b} h^2$ &$0.02240^{+0.00020}_{-0.00020}$&$0.02251^{+0.00021}_{-0.00026}$&$0.022413\pm 0.000088$\\
			$Y_p$  &$0.24682^{+0.00073}_{-0.0015}$&$0.24681^{+0.00079}_{-0.0015}$&$0.24638^{+0.00042}_{-0.00064}$\\
			$10^{5}\cdot(D/H)$ &$2.547^{+0.030}_{-0.041}$&$2.527\pm 0.030$&$2.533\pm 0.019$\\
			$\Delta N_{\rm eff}$ &$< 0.31$ \,\,$(< 0.40)$&$< 0.30\,\,(<0.40)$&$< 0.14\,\,(< 0.21)$\\
			\hline
			$m_a$ [eV] &$< 0.61 \,\, (< 3.6)$&$<0.53\,\, (<3.0  )$&$< 0.16\,\,(<0.25)$\\
			$f_a$ [$10^7$ GeV]&$>0.934\, (>0.158)$&$>1.07\,(>0.19)$&$>3.56\,(>2.28)$\\
			\hline	\hline
		\multicolumn{4}{c}{\textbf{DSFZ Axion Model ($\tan \beta=3$)}}\\
		    \hline\hline
			$\Omega_{\rm b} h^2$ &$0.02240\pm 0.00010$&$0.02249^{+0.00022}_{-0.00027}$&$0.022407\pm 0.000089$\\
			$Y_p$  &$0.24668^{+0.00075}_{-0.0016}$&$0.24665^{+0.00082}_{-0.0016}$&$0.24621^{+0.00045}_{-0.00068} $\\
			$10^{5}\cdot(D/H)$ &$2.543^{+0.031}_{-0.044}$&$2.526\pm 0.030$&$2.530\pm 0.019$\\
			$\Delta N_{\rm eff}$ &$< 0.32\,\,(< 0.41)$&$< 0.30\,\,(< 0.41)$&$< 0.14\,\,(< 0.20)$\\
			\hline
			$m_a$ [eV] &$< 0.40\,\,(< 2.4)$&$< 0.33\,\,(< 2.0)$&$< 0.10\,\,(< 0.15)$\\
			$f_a$ [$10^7$ GeV] &$>1.42\,(>0.23)$&$>1.73\,(>0.28)$&$>5.7\,(>3.8)$\\
			\hline	\hline
		\multicolumn{4}{c}{\textbf{DSFZ Axion Model ($\tan \beta=10$)}}\\
		    \hline\hline
			$\Omega_{\rm b} h^2$ &$0.02240\pm 0.00010$&$0.02249^{+0.00022}_{-0.00027}$&$0.022407\pm 0.000089$\\
			$Y_p$  &$0.24671^{+0.00077}_{-0.0017}$&$0.24668^{+0.00086}_{-0.0017}$&$0.24622^{+0.00049}_{-0.00076}$\\
			$10^{5}\cdot(D/H)$ &$2.544^{+0.032}_{-0.045}$&$2.526\pm 0.030$&$2.530\pm 0.020$\\
			$\Delta N_{\rm eff}$ &$< 0.33\,\,(< 0.42)$&$< 0.31\,\,(< 0.41)$&$< 0.14\,\,(< 0.21)$\\
			\hline
			$m_a$ [eV] &$< 0.40\,\,(< 2.5)$&$< 0.33\,\,(< 2.0)$&$< 0.10\,\,(< 0.15)$\\
			$f_a$ [$10^7$ GeV] &$>1.42\,(>0.23)$&$>1.73\,(>0.28)$&$>5.7\,(>3.8)$\\
			\hline	\hline
		\end{tabular}}
	\end{center}
	\caption{Results obtained from BBN primordial abundances. The constraints on $\Omega_{\rm b} h^2$ , $Y_p$ and $10^{5}\cdot(D/H)$ are given at 68\%CL while the upper bounds on $\Delta N_{\rm eff}$ and $m_a$ are given both at 95\% and 99\% CL. The horizontal lines divide the fit parameters from the derived parameters.}
	\label{tab.BBN}
\end{table*}

\begin{figure*}
\centering
\includegraphics[width=\textwidth]{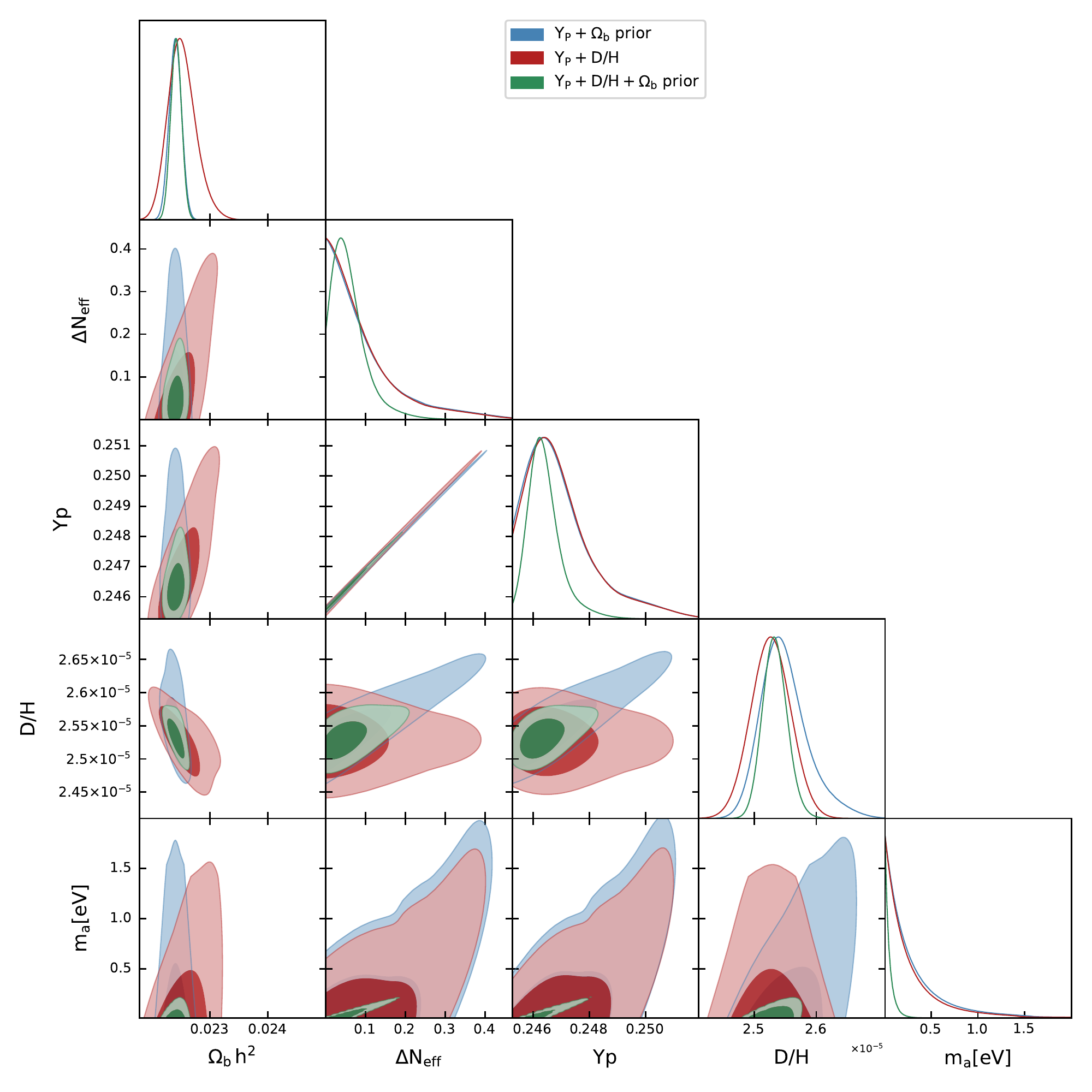}
\caption{Two-dimensional 68\% and 95\% CL allowed regions and one-dimensional probability posterior distributions for the most relevant cosmological parameters in axion cosmologies within the KSVZ QCD axion scenario. The different colors refer to the different data combinations here considered for BBN analyses, see \autoref{tab.BBN}.}
\label{fig:KSVZ_BBN}
\end{figure*}

\begin{figure*}
\centering
\includegraphics[width=\textwidth]{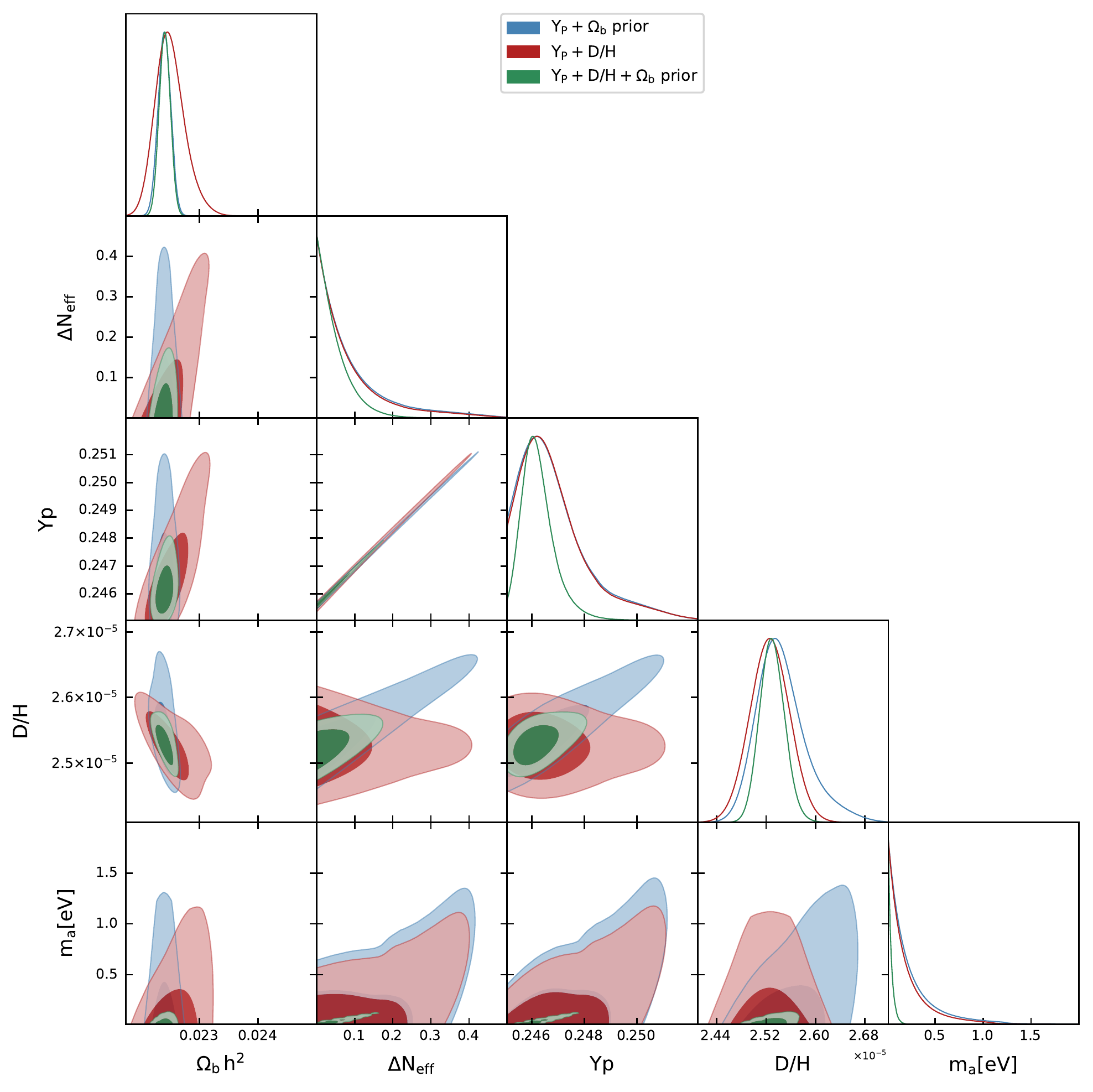}
\caption{As \autoref{fig:KSVZ_BBN} but for the DFSZ QCD axion scenario. Since the results shown in Tab.~\ref{tab.BBN} demonstrate that the constraints in this axion cosmological scenario are almost independent of the $\beta$ parameter, we only illustrate here the case $\tan \beta=3$.}
\label{fig:DFSZ_BBN}
\end{figure*}

It is instructive to start studying the constraints on thermal QCD axions restricting the analysis only to the BBN epoch. As explained in the previous section, the only effect of the axion during BBN is its contribution to the radiation energy density of the Universe via $\Delta N_{\rm eff}$. Independent astrophysical and cosmological measurements of primordial light element abundances limit the amount of additional dark radiation and thus impose an upper bound on the mass of the axion. Our baseline dataset for the BBN analyses consists of 
\begin{itemize}
\item Two independent measurements of the primordial Helium \textit{mass fraction}, $Y_p = 0.2449 \pm 0.0040$ \cite{Aver:2015iza} and $Y_p = 0.2446\pm 0.0029$ \cite{Peimbert:2016bdg}. 

\item A percent determination of the primordial Deuterium abundance $D/H=\left(2.527\pm0.030\right)\cdot 10^{-5}$ based on six high precision and homogeneously analyzed $D/H$ measurements from \cite{Cooke:2017cwo}.

\item The value of the baryon energy density parameter $\Omega_b\,h^2=0.0224 \pm 0.0001$ from the final 2018 Planck data release of temperature and polarization CMB angular power spectra \cite{Aghanim:2018eyx}.

\end{itemize}

We randomly sample $N=10^6$ linearly distributed values of the axion mass in the range $m_a\in[0.1\,,\,10]$ eV and of the baryon energy density in the range $\Omega_b h^2\in\left[0.020\,,\,0.025 \right]$. For each value of the axion mass, we compute the axion contribution to the effective number of neutrino species $\Delta N_{\rm eff} (m_a)$ by solving the Boltzmann equation. Given a set of points $(\Delta N_{\rm eff} ,\Omega_b h^2)$, we compute the BBN predictions for the primordial light elements by means of the publicly available code \textsc{PArthENoPE}, following the procedure outlined in the previous section. We therefore create a grid of points in the $(\Delta N_{\rm eff} ,\Omega_b h^2)$ plane similar to those usually obtained within the Monte Carlo methods. We then apply the priors on the BBN abundances, re-weighting the contributions of the points by means of an ``importance sampling'' method. As a result, we obtain informative posterior distributions for the parameters to be inferred by observations such as the axion mass and $\Delta N_{\rm eff}$. We apply this methodology to the KSVZ and DFSZ axion models. \autoref{tab.BBN} summarizes our results, \autoref{fig:KSVZ_BBN} and \autoref{fig:DFSZ_BBN} provide the marginalized posterior distributions of parameters for the KSVZ and the DFSZ axion models, respectively. For the latter model, we fix the additional free parameter $\beta$ to two different values, $\tan \beta= \{3, 10\}$.  

As a common practice, we start adopting prior information on the total amount of the primordial Helium $Y_p$ and the baryon energy density $\Omega_b\,h^2$. We refer to this dataset as ``BBN-A ($Y_p + \Omega_b\,h^2$)'', which provides an upper limit on the additional dark radiation allowed during BBN $\Delta N_{\rm eff}<0.3$ at 95\% CL ( $\Delta N_{\rm eff}<0.4$ at 99\% CL), in perfect agreement with previous results discussed in literature~\cite{Giare:2020vzo}. Notice also that this bound remains unchanged for both models of axions, as expected. For the KSVZ scenario, this limit is translated into an upper bound on the axion mass of $m_a<0.6$ eV at 95\%CL ($m_a<3.6$ eV at 99\%CL). The axion contribution to $\Delta N_{\rm eff}$ is typically larger within the DFSZ model, and fixing $\tan \beta=3$ we obtain a slightly more constraining bound $m_a<0.4$ eV at 95\%CL ($m_a<2.4$ eV at 99\%CL). Performing the very same analysis with $\tan \beta=10$ the result does not change, proving that differences between the two cases remain too small to be observed given the current uncertainties, see also \autoref{fig:DFSZ_BBN} and \autoref{tab.BBN}. Finally, the constraints on the total amount of primordial Deuterium are at the percent level $D/H\simeq 2.54^{+0.03}_{-0.04}\cdot 10^{-5}$ (at 68\% CL) for both KSVZ and DFSZ models. Since in the analysis labeled here as BBN-A we are not assuming any prior information on Deuterium, this result can be compared to direct astrophysical measurements ($D/H\simeq 2.527\pm 0.030 \cdot 10^{-5}$), showing a perfect agreement and providing a robust consistency check. 

Notice however that these results are derived assuming a prior knowledge of the baryon-energy density as inferred by the Plank collaboration analyzing the last release of the CMB data. In order to remain completely independent on the physics at the recombination epoch, we repeat the same analysis by relaxing the prior on $\Omega_b h^2$ and combining instead with information on the total amount of both primordial Helium $Y_p$ and Deuterium $D/H$ from direct astrophysical measurements. We refer to this dataset as ``BBN-B($Y_p + D/H$)''. Also in this case, we derive exactly the same limit $\Delta N_{\rm eff}<0.3$ at 95\%CL ($\Delta N_{\rm eff}<0.4$ at 99\% CL) on additional dark radiation at the BBN epoch. However, the bound on the axion mass slightly changes with respect to the case BBN-A because of the different correlations among the parameters. In particular, assuming the prior on $D/H$ breaks the strong positive correlation between primordial Deuterium and the axion mass present in the previous case, see also \autoref{fig:KSVZ_BBN} and \autoref{fig:DFSZ_BBN}. This leads to a small improvement in the axion mass limit: $m_a<0.5$ eV at 95\%CL ($m_a<3.0$ eV at 99\%CL) for the KSVZ axion, and $m_a<0.3$ eV at 95\%CL ($m_a<2.0$ eV at 99\%CL) for the DFSZ model (again with no differences between $\tan \beta=3$ and $\tan\beta=10$). We are not imposing any prior information on the baryon mass-energy density, which turns out to be $\Omega_b\,h^2=0.0225^{+0.0002}_{-0.0003}$ both for the KSVZ and DFSZ models. Since this bounds is derived only by assuming a prior knowledge of primordial abundances of Helium and Deuterium, it is completely independent from the CMB and thus can be compared with the value inferred by Planck collaboration ($\Omega_b\,h^2=0.0224 \pm 0.0001$), showing again a perfect agreement and providing another important consistency check of our numerical analyses. 

Finally, for completeness, we combine all the three priors together, referring to this dataset as ``BBN-C($Y_p+D/H+\Omega_b\,h^2$)''. This combination provides an improvement in the constraining power on the different parameters, leading to the limit $\Delta N_{\rm eff}<0.14$ at 95\% CL ($\Delta N_{\rm eff}<0.2$ at 99\% CL). This is a factor $\sim 2$ stronger than the bound from the previous two datasets (BBN-A and BBN-B). Consequently, the limits on the axion mass are also improved, obtaining in this case $m_a<0.16$ eV at 95\% CL ($m_a<0.25$~eV at 99\% CL) for the KSVZ  model and $m_a<0.10$ eV at 95\% CL ($m_a<0.15$~eV at 99\% CL) for the DFSZ scenario. While this improvement is of course interesting to be noticed, it is also worth stressing that the combination of three different priors from the BBN perspective is not needed, as only two input parameters (\textit{i.e.}, $\Delta N_{\rm eff}$ and $\Omega_b\,h^2$) are required to compute all the light element abundances. Nevertheless, the additional constraints do not produce shifts or biases in the results, and only reduce the uncertainties, see also \autoref{fig:KSVZ_BBN} and \autoref{fig:DFSZ_BBN}, reassessing the consistency among the data sets used here, which show no tensions, and further confirming the robustness of our analyses. 

\subsection{Results from the full cosmological analyses}

The previous results have shown that the axion contribution to the radiation energy density is severely constrained during the BBN epoch both by direct astrophysical measurements of primordial light abundances and by constraints on the baryon energy density, limiting the allowed axion masses to the sub-eV range. Nevertheless the analyses carried out so far focused exclusively on a precise cosmological era, ignoring the subsequent Universe evolution. We shall perform in what follows a comprehensive analysis, addressing all the axion contributions towards the different cosmological epochs/scales. 

The fiducial cosmological model that we analyze is a \emph{hot-relic} extension of the $\Lambda$CDM model, including both axions and neutrinos as hot thermal massive relics. Therefore, in addition to the $\Lambda$CDM six-parameter model (\emph{i.e.}, the baryon $\omega_{\rm b}\equiv \Omega_{\rm b}h^2$ and cold dark matter $\omega_{\rm c}\equiv\Omega_{\rm c}h^2$ energy densities, the angular size of the horizon at the last scattering surface $\theta_{\rm{MC}}$, the optical depth $\tau$, the amplitude of primordial scalar perturbation $\log(10^{10}A_{\rm S})$ and the scalar spectral index $n_{\rm S}$), we consider also the axion mass $m_a$ and the sum of three active neutrino masses $\sum m_{\nu}$ (both in $\rm{eV}$). Concerning the QCD axion, we carry out two separate studies for the KSVZ and DFSZ frameworks \footnote{The previous BBN analyses have shown that changes in the $\beta$ parameter are negligible. Therefore, in the full MCMC analysis within the DFSZ scenario we shall fix $\tan \beta=3$.} and, in both cases, we perform Monte Carlo Markov Chain (MCMC) analyses using the publicly available version of the \textsc{COBAYA} software~\citep{Torrado:2020xyz} and computing the theoretical model with the modified version of the cosmological Boltzmann integrator code \textsc{CAMB} \citep{Lewis:1999bs,Howlett:2012mh} described in \autoref{sec.Cosmo.Implementation}. We choose flat prior-distributions for all the above mentioned cosmological parameters (unless otherwise stated), varying them uniformly in the conservative ranges listed in \autoref{tab.Priors}. The convergence of the chains obtained with this procedure is tested using the Gelman-Rubin criterion~\cite{Gelman:1992zz} and we choose as a threshold for chain convergence $R-1 \lesssim 0.02 $. Finally, we explore the posteriors of our parameter space using the sampler developed for \texttt{CosmoMC}~\cite{Lewis:2002ah,Lewis:2013hha}, see also Ref.~\cite{Neal:2005} for details concerning the ``fast dragging'' procedure used in our sampling method.

Concerning observational constraints, our baseline data-set includes:
\begin{itemize}
	
	\item Planck 2018 temperature and polarization (TT TE EE) likelihood, which also includes low multipole data ($\ell < 30$)~\citep{Aghanim:2019ame,Aghanim:2018eyx,Akrami:2018vks}. We refer to this combination as ``Planck 2018''.
	
	\item Planck 2018 lensing likelihood~\citep{Aghanim:2018oex}, constructed from measurements of the power spectrum of the lensing potential. We refer to this dataset as ``lensing''.
	
	\item Baryon Acoustic Oscillations (BAO) measurements extracted from data from the 6dFGS~\cite{Beutler:2011hx}, SDSS MGS~\cite{Ross:2014qpa} and BOSS DR12~\cite{Alam:2016hwk} surveys. We refer to this dataset combination as ``BAO''.
	
\end{itemize}

\label{sec.Results.CMB}
\begin{table}
	\begin{center}
		\renewcommand{\arraystretch}{1.5}
		\begin{tabular}{l@{\hspace{0. cm}}@{\hspace{1.5 cm}} c}
			\hline
			\textbf{Parameter}    & \textbf{Prior} \\
			\hline\hline
			$\Omega_{\rm b} h^2$         & $[0.005\,,\,0.  1]$ \\
			$\Omega_{\rm c} h^2$     	 & $[0.001\,,\,0.99]$\\
			$100\,\theta_{\rm {MC}}$     & $[0.5\,,\,10]$ \\
			$\tau$                       & $[0.01\,,\,0.8]$\\
			$\log(10^{10}A_{\rm S})$     & $[1.61\,,\,3.91]$ \\
			$n_{\rm s}$                  & $[0.8\,,\, 1.2]$ \\
			$\sum m_{\nu}$ [eV]          & $[0.06\,,\,5]$\\
			$m_a$ [eV]                   & $[0.1\,,\,10]$\\
			\hline\hline
		\end{tabular}
		\caption{List of the parameter priors.}
		\label{tab.Priors}
	\end{center}
\end{table}

\begin{table*}[htpb!]
	\begin{center}
		\renewcommand{\arraystretch}{1.5}
		\resizebox{0.7\textwidth}{!}{\begin{tabular}{l c c c}
		\multicolumn{4}{c}{}\\
		\multicolumn{4}{c}{\textbf{KSVZ Axion Model}}\\[0.2cm]
  	        \hline
		    \textbf{Parameter} & \textbf{\nq{Planck 2018\\ (TT TE EE) }}  & \textbf{\nq{Planck 2018\\+ lensing}}  &  \textbf{\nq{Planck 2018\\+lensing+BAO}} \\
			\hline\hline
			$\Omega_{\rm b} h^2$ &$0.02244\pm 0.00017$&$0.02243\pm 0.00015$&$0.02250\pm 0.00015$\\
			$\Omega_{\rm c} h^2$ &$0.1224^{+0.0018}_{-0.0021}$&$0.1225^{+0.0016}_{-0.0022}$&$0.1208^{+0.0012}_{-0.0014}$\\
			$100\,\theta_{\rm {MC}}$ &$1.04060^{+0.00038}_{-0.00034}$&$1.04058\pm 0.00036$&$1.04081\pm 0.00032$\\
			$\tau$   &$0.0557\pm 0.0079$&$0.0564\pm 0.0076$&$0.0581^{+0.0070}_{-0.0080}$\\
			$\log(10^{10}A_{\rm S})$ &$3.053\pm 0.017 $&$3.055^{+0.015}_{-0.016}$&$3.055^{+0.014}_{-0.016}$\\
			$n_s$ &$0.9687\pm 0.0051$&$0.9681\pm 0.0049$&$0.9703\pm 0.0043$\\
			$m_a$ [eV] &$< 1.04\,\,(< 1.86)$&$< 0.888\,\,(< 1.67)$&$< 0.282\,\,(< 0.420)$\\
			$\sum m_{\nu}$ [eV] &$< 0.297\,\,(< 0.422)$&$< 0.278\,\,(< 0.381)$&$<0.156\,\,(< 0.192)$\\
			\hline
			$f_a$ [$10^{7}$ GeV]&$>0.55\,(>0.31)$&$>0.64\,(>0.34)$&$>2.02\,(>1.35)$\\
			$\Delta N_{\rm eff}$ &$< 0.349\,\,(< 0.378)$&$< 0.340\,\,(<0.373)$&$< 0.226\,\,(< 0.275)$\\
			$Y_P$&$0.24746^{+0.00077}_{-0.0017} $&$0.24741^{+0.00074}_{-0.0016}$&$0.24693^{+0.00048}_{-0.00097}$\\
			$10^5\cdot (D/H)$&$2.556^{+0.033}_{-0.042}$&$2.558^{+0.029}_{-0.040}$&$2.531\pm 0.028$\\
			$H_0$ [Km/s/Mpc] &$67.0^{+1.2}_{-0.77}$&$66.9^{+1.2}_{-0.73}$&$67.90\pm 0.53$\\
			$\sigma_8$ &$0.790^{+0.029}_{-0.012}$&$0.793^{+0.023}_{-0.011}$&$0.8052^{+0.0099}_{-0.0075}$\\
			\hline	\hline
		\end{tabular}}
	\end{center}
	\caption{Results obtained for the KSVZ Axion Model. The constraints on parameters are given at 68\%CL while the upper bounds are given both at 95\% and at 99\% CL. The horizontal line divides the fit parameters of the cosmological model from the derived ones.}
	\label{tab.CMB.KSVZ}
\end{table*}

\begin{table*}[htpb!]
	\begin{center}
		\renewcommand{\arraystretch}{1.5}
		\resizebox{0.7\textwidth}{!}{\begin{tabular}{l c c c}
		\multicolumn{4}{c}{}\\
		\multicolumn{4}{c}{\textbf{DSFZ Axion Model ($\tan \beta=3$)}}\\[0.2cm]
  	        \hline			
		    \textbf{Parameter} & \textbf{\nq{Planck 2018\\ (TT TE EE) }}  & \textbf{\nq{Planck 2018\\+ lensing}}  &  \textbf{\nq{Planck 2018\\+lensing+BAO}} \\
			\hline\hline
			$\Omega_{\rm b} h^2$ &$0.02243\pm 0.00017$&$0.02244\pm 0.00017$&$0.02250\pm 0.00015$\\
			$\Omega_{\rm c} h^2$ &$0.1223^{+0.0018}_{-0.0023}$&$ 0.1224^{+0.0017}_{-0.0024}$&$0.1208^{+0.0012}_{-0.0016}$\\
			$100\,\theta_{\rm {MC}}$ &$1.04061^{+0.00039}_{-0.00035}$&$1.04061^{+0.00040}_{-0.00035}$&$1.04081\pm 0.00031$\\
			$\tau$   &$ 0.0554\pm 0.0080 $&$0.0566^{+0.0071}_{-0.0081}$&$0.0576^{+0.0069}_{-0.0077}$\\
			$\log(10^{10}A_{\rm S})$&$3.052\pm 0.017$&$3.055^{+0.015}_{-0.017}$&$3.054\pm 0.015$\\
			$n_s$ &$0.9685\pm 0.0053 $&$0.9682\pm 0.0052 $&$0.9702^{+0.0042}_{-0.0048}$\\
			$m_a$ [eV] &$< 0.710 \, (< 1.81)$&$ < 0.697  \, (< 1.42)$&$< 0.209 \, (< 0.293)$\\
			$\sum m_{\nu}$ [eV] &$< 0.312 \, (< 0.426)$&$< 0.288 \,  (< 0.390)$&$< 0.157 \, (< 0.193)$\\
			\hline
			$f_a$ [$10^7$ GeV] &$>0.80\,(>0.31)$&$>0.82\,(>0.40)$&$>2.72\,(>1.94)$\\
			$\Delta N_{\rm eff}$ &$< 0.358 \, (< 0.401 ) $&$< 0.360\, (<0.392)$&$< 0.243\,(<0.290)$\\
			$Y_P$&$0.24732^{+0.00077}_{-0.0018}$&$0.24731^{+0.00077}_{-0.0018}$&$0.24689^{+0.00061}_{-0.0012}$\\
			$10^5\cdot (D/H)$&$2.554^{+0.032}_{-0.041}$&$2.554^{+0.032}_{-0.042}$&$2.531\pm 0.028$\\
			$H_0$ [Km/s/Mpc] &$67.0^{+1.2}_{-0.82}$&$67.0^{+1.2}_{-0.79} $&$67.90\pm 0.58$\\
			$\sigma_8$ &$0.792^{+0.027}_{-0.012}$&$0.794^{+0.022}_{-0.010}$&$0.8059^{+0.0093}_{-0.0071}$\\
			\hline	\hline
		\end{tabular}}
	\end{center}
	\caption{Results obtained for the DFSZ Axion Model ($\tan \beta=3$). The constraints on parameters are given at 68\%CL while the upper bounds are given both at 95\% and at 99\% CL. The horizontal line divides the fit parameters of the cosmological model from the derived ones.}
	\label{tab.CMB.DFSZ}
\end{table*}

\begin{figure*}[htpb!]
\centering
\includegraphics[width=\textwidth]{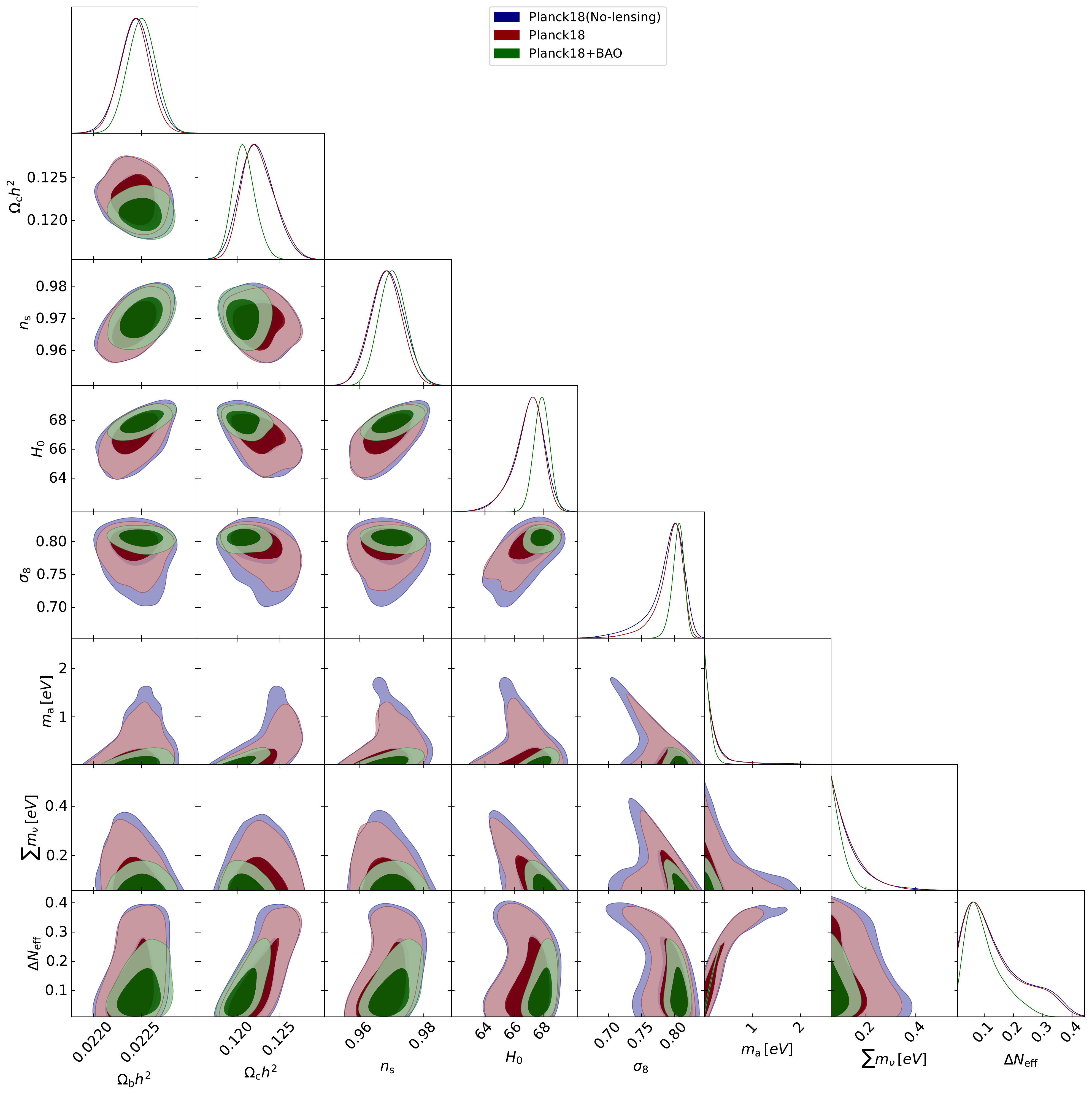}
\caption{Two-dimensional 68\% and 95\% CL allowed regions and one-dimensional probability posterior distributions for the most relevant cosmological parameters in axion cosmologies within the KSVZ QCD axion scenario. The different colors refer to the different data combinations considered for MCMC analyses, see \autoref{tab.CMB.KSVZ}.}
\label{fig:Planck_KSVZ}
\end{figure*}

\begin{figure*}[htpb!]
\centering
\includegraphics[width=\textwidth]{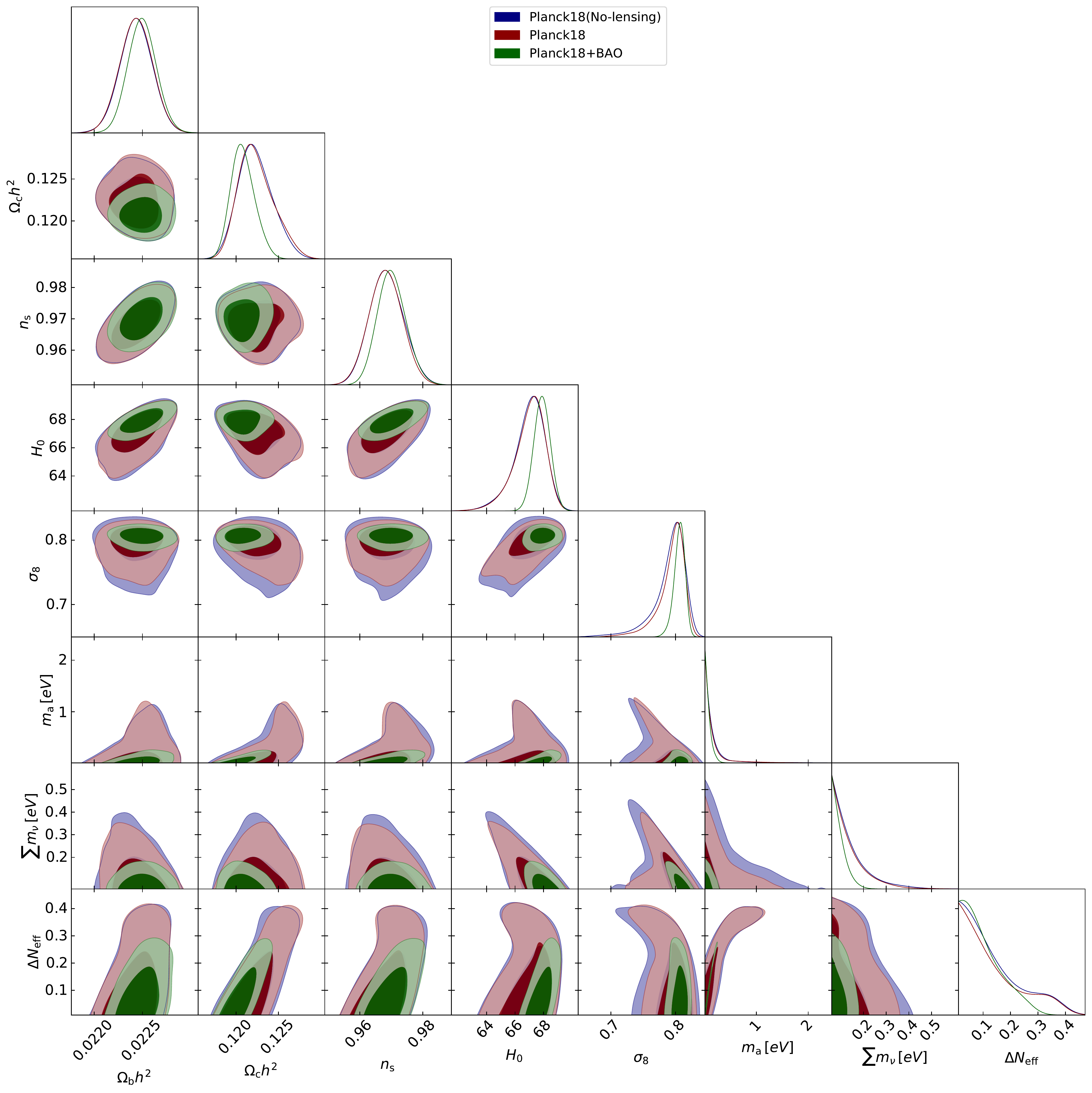}
\caption{Two-dimensional 68\% and 95\% CL allowed regions and one-dimensional probability posterior distributions for the most relevant cosmological parameters in axion cosmologies within the DFSZ QCD axion scenario with $\tan \beta=3$. The different colors refer to the different data combinations considered for MCMC analyses, see \autoref{tab.CMB.DFSZ}.}
\label{fig:Planck_DSFZ}
\end{figure*}

\begin{figure*}[htpb!]
\centering
\includegraphics[width=\textwidth]{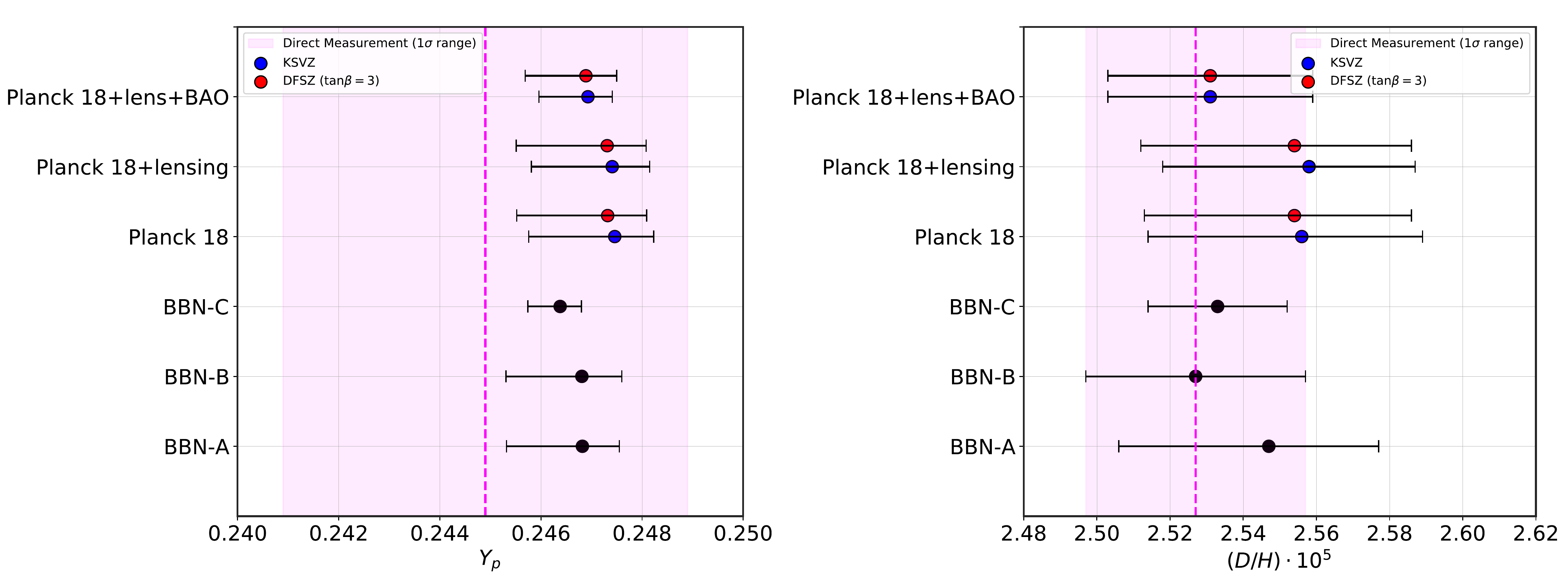}
\caption{Constraints on the primordial abundance of Helium-4 ($Y_p$, left panel) and Deuterium ($D/H$, right panel) derived within the different methods and data-set detailed in the manuscript. The pink bands identify $68\%$ C.L. from current astrophysical determinations of $Y_p$ \cite{Aver:2015iza} and $D/H$ \cite{Cooke:2017cwo}.}
\label{fig:Elements}
\end{figure*}

In particular, combining the Planck (TT TE EE) spectra with the Planck lensing measurement can be useful to gain some additional sensitivity on parameters that affect both the late-time expansion and the background geometry while the inclusion of the large-scale structure information from BAO measurements from galaxy surveys turns out to become crucial to set bounds on the amount of dark matter in the form of hot relics, as axions and/or neutrinos, as it is a powerful way to break degeneracy in the geometrical sector. 

We summarize the constraints on the cosmological parameters (at 68\% CL) and the upper bounds on the axion and neutrino masses (both at 95\% and 99\% CL) in \autoref{tab.CMB.KSVZ} and \autoref{tab.CMB.DFSZ} for the KSVZ and DFSZ model, respectively. We also show the corresponding 1D and 2D marginalized posteriors in \autoref{fig:Planck_KSVZ} and in \autoref{fig:Planck_DSFZ}.

As concerns the KSVZ axion model, exploiting the 2018 release of Planck temperature and polarization (TT TE EE) data, we derive the upper bound $m_a<1.04\,\rm{eV}$ at 95\% CL ($m_a<1.86\,\rm{eV}$ at 99\% CL) on the axion mass and $\sum m_{\nu}<0.297\,\rm{eV}$ at 95\% CL ($\sum m_{\nu}<0.422\,\rm{eV}$ at 99\% CL) on the sum of the neutrino masses. Notice that now the 95\% CL bound on the axion mass is much less constraining than those previously derived in \autoref{sec.Results.BBN}. Indeed, in this case we have many more parameters and therefore way more degeneracies, especially for the case of the neutrino mass and the cold dark matter energy density. Additionally, we include here the BBN predictions in the MCMC analysis by relating $\Omega_b$ and $Y_P$ through \autoref{eq.Yp} and thus without including direct astrophysical measurements of primordial abundances. While we obtain a slightly less constraining upper limit (even if perfectly consistent with the previous bounds from BBN light element abundances) for the total amount of extra dark radiation due to the axion contribution, $\Delta N_{\rm eff}<0.349$ at 95\% CL, the CMB analyses are fully consistent with both the standard BBN predictions and  direct astrophysical measurements, see also \autoref{tab.CMB.KSVZ} and \autoref{fig:Elements} where we compare the values of $Y_P$ and $D/H$ as inferred by the different analysis and datasets. In particular, \autoref{fig:Elements} clearly states the agreement from the different data combinations considered along this study in the extraction of primordial element abundances,  reassessing the limits on the different cosmological parameters derived here.

Interestingly, the 99\% CL limits on the axion mass turn out to be more constrained than the BBN 99\% CL bounds, especially when including large scale structure observations. This is due to the fact that the large masses allowed within the 99\% CL contours during BBN are now excluded by their late-time effects on structure formation (providing also evidence about the importance of extending the analysis to all the different cosmological epochs and scales). 

As concerns the other datasets considered in our analyses, we may appreciate that the inclusion of CMB lensing measurements from the Planck satellite improve both the KSVZ axion mass bound $m_a<0.888$ eV at 95\% CL ($m_a<1.67$ eV at 99\% CL) and the limit on neutrinos $\sum m_{\nu}<0.278$ eV at 95\% CL ($\sum m_a<0.381$ eV at 99\% CL). However, as aforementioned, the most significant improvement comes from the inclusion of BAO data. QCD thermal axions and neutrinos are hot thermal relics with very large dispersion velocities, erasing structure formation below their free streaming scale, directly related to their mass. Therefore, none of these hot thermal species will contribute to structure formation at small scales, being the suppression of the matter power spectrum in the linear regime proportional to the amount of dark matter in the form of hot relics. Consequently, the addition of BAO observations leads to a robust upper limit of $m_a<0.282$ eV at 95\% CL ($m_a<0.420$ eV at 99\% CL) simultaneously improving the limit on the total neutrino mass to $\sum m_{\nu}<0.156$ eV at 95\% CL ($\sum m_a<0.192$ eV at 99\% CL). 

The same analyses for the DFSZ scenario leads to bounds similar to those obtained for the KSVZ axion (as in the BBN case of the previous section). In particular, analyzing the Planck (TT TE EE) data we obtain a more constraining bound on the axion mass with respect to the KSVZ scenario, namely: $m_a<0.710$ eV at 95\% CL ($m_a<1.81$ eV at 99\% CL). Including both lensing and BAO these limits are improved to $m_a<0.209$ eV at 95\% CL ($m_a<0.293$ eV at 99\% CL). Conversely, the limits on the total neutrino mass remain basically unchanged.

Finally, we stress that the bounds on the axion mass obtained in this work are a factor $\sim 5$ better with respect to the results derived in Ref.~\cite{Giare:2020vzo} exploiting the very same data-sets.  Indeed, in that case the analysis was performed for the axion-pion and the axion-gluon couplings separately. For the former, it was only considered the parameter space where the chiral perturbation theory approach was reliable (\textit{i.e.}, to decoupling temperatures smaller than $62$ MeV). Here, we have overcome both problems smoothly extending the thermal axion production across the QCD phase transition. Furthermore, we improve our predictions for the amount of axions by solving the Boltzmann equation instead of relying upon the instantaneous decoupling approximation.

\subsection{A Forecast for CMB-S4 and DESI}

One of the key targets of the Stage-IV CMB experiment (CMB-S4) is to increase the accuracy on extra dark radiation by almost an order of magnitude, opening to the possibility of robustly constraining the mass of thermal relics such as  QCD axions~ \cite{Abazajian:2016yjj}. In addition, future observations of large scale structure will lead to highly accurate Baryon Acoustic Oscillation (BAO) data and are expected to provide an enormous improvement on the neutrino mass bound~\cite{DESI}. Here we conclude our analyses studying  how  these future experimental efforts will improve the constraining power on hot relics discussed so far. In particular, we simulate data for a CMB-S4-like \cite{Abazajian:2016yjj} observatory and for a DESI-like \cite{DESI,DESI:2013agm} BAO galaxy survey. These two future probes are expected to provide scientific results in the next few years and have been carefully designed to significantly improve the constraints on the neutrino sector and other forms of dark radiation, see e.g. Refs.~\cite{Abazajian:2016yjj,DESI,constraint_neutrino_stageIV,Abazajian:2019oqj}.

\begin{table*}
	\begin{center}
		\renewcommand{\arraystretch}{1.5}
		\resizebox{0.7 \textwidth}{!}{\begin{tabular}{l c c c }
		\multicolumn{4}{c}{}\\
		\multicolumn{4}{c}{\textbf{Forecast $\Lambda$CDM +$m_a=1$eV}}\\[0.2cm]
  	        \hline
		    \textbf{Parameter} & \textbf{Fiducial Value} & \textbf{CMB-S4}  & \textbf{CMB-S4 + DESI} \\
			\hline\hline
			$\Omega_{\rm b} h^2$ &$0.0224$&$0.022399\pm 0.000035$&$0.022399\pm 0.000034$\\
			$\Omega_{\rm c} h^2$ &$0.12$&$0.12050\pm 0.00059$&$0.11984\pm 0.00031$\\
		    $H_0$ [Km/s/Mpc]&$67.4$&$66.57^{+0.57}_{-0.44}$&$67.16\pm 0.22 $\\
			$\tau$   &$0.05$&$0.0506^{+0.0024}_{-0.0026}$&$0.0511^{+0.0024}_{-0.0027}$\\
			$\log(10^{10}A_{\rm S})$ &$3.044$&$3.0484\pm 0.0047$&$3.0475\pm 0.0047$\\
			$n_s$ &$0.965$&$0.9622\pm 0.0021$&$0.9639\pm 0.0017$\\
			$m_a$ [eV] &$1$&$1.13\pm 0.11$&$1.056^{+0.086}_{-0.075}$\\
			$\sum m_{\nu}$ [eV] &$0.06$&$< 0.178\,\,(<0.220)$&$< 0.118\,\,(< 0.144)$\\
			\hline
			$f_a$ [$10^7$ GeV] &$0.570$ &$0.519^{+0.040}_{-0.056}$&$0.543^{+0.033}_{-0.048}$\\
			$\Delta N_{\rm eff}$ &$0.347$&$0.3535^{+0.0060}_{-0.0044}$&$0.3497^{+0.0051}_{-0.0036}$\\
			$Y_P$&$0.250114$&$0.250135^{+0.000079}_{-0.000070}$ & $0.250087^{+0.000068}_{-0.000049}$\\
			$10^5\cdot (D/H)$&$2.6326$&$2.6350\pm 0.0068$&$2.6338\pm 0.0066$\\
			$\sigma_8$&$0.7323$&$0.7191^{+0.0059}_{-0.0050}$&$0.7255\pm 0.0028$\\
			\hline	\hline
		\end{tabular}}
	\end{center}
	\caption{Forecasted mean values, errors and upper bounds on the different cosmological parameters rising from  CMB-S4 only and from CMB-S4 plus DESI. The forecasted errors on the parameters are given at 68\%CL, while the upper limits are given both at 95\% and at 99\% CL. The horizontal line divides the fit parameters of the cosmological model from the derived ones.}
	\label{tab.forecast}
\end{table*}

\begin{figure*}
\centering
\includegraphics[width=\textwidth]{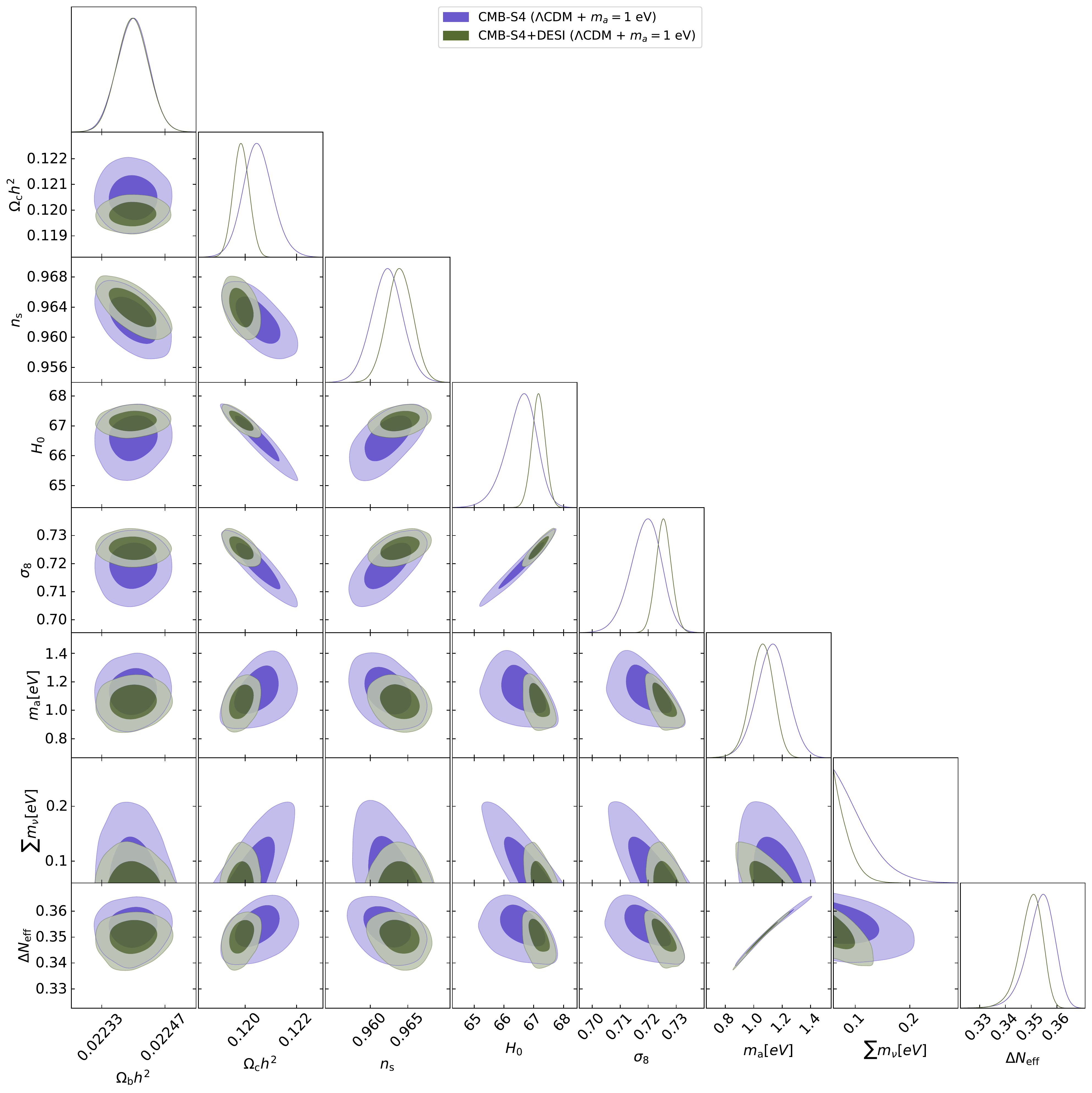}
\caption{Two-dimensional 68\% and 95\% CL allowed regions and one-dimensional probability posterior distributions for the most relevant cosmological parameters in axion cosmologies obtained from the forecasting data and methods, see \autoref{tab.forecast}.}
\label{fig:forecast}
\end{figure*}

The fiducial cosmological model used to build the forecasted data is a \emph{hot-relic} extension of the $\Lambda$CDM model that assumes a non-vanishing axion mass ($m_a=1$ eV) and a non-vanishing total neutrino mass $\sum m_\nu = 0.06$ eV i.e. the minimum mass allowed by neutrino oscillation data. We adopt the KSVZ model for governing the axion interactions. The other standard six cosmological parameters are chosen to be in agreement wit the latest Planck 2018 constraints for a $\Lambda$CDM scenario, namely, we adopt the following fiducial values:  $n_s = 0.965$, $\omega_b = 0.0224$, $\omega_c = 0.12$, $H_0 = 67.4$, $\tau=0.05$, $A_s = 2.1 \times 10^{-9}$. Notice that, within the KSVZ scenario, values of the axion mass $\sim 1$ eV, are still marginally consistent with the 95\% CL limit derived from the CMB data but they are strongly disfavored by large scale structure measurements from BAO. Therefore, one would expect to recover the value of axion mass assumed in our fiducial model with an uncertainty $\sigma(m_a)\ll 1\,\rm{eV}$,  that will remark the sensitivity on the axion mass from future cosmic probes.

In order to generate the forecasted data, we followed the same procedure detailed in \cite{Giare:2021cqr}:  using the fiducial model introduced above, we compute the angular power spectra of temperature $ C_\ell^{TT} $, E and B polarization $ C_\ell^{EE,BB} $ and cross temperature-polarization $C_\ell^{TE} $ anisotropies. Then, we consider an experimental noise for the temperature angular spectra of the form \citep{Perotto:2006rj}
\begin{equation}
	N_\ell = w^{-1}\exp(\ell(\ell+1)\theta^2/8 \ln 2)~,
\end{equation}  
where $ \theta $ is the FWHM angular resolution and $ w^{-1} $ is the experimental sensitivity in units of $ \mu\mathrm{K}\,\rm arcmin $. The polarization noise is derived assuming $ w_p^{-1} = 2w^{-1} $ (one detector measures two polarization states). Here we fix $ \theta = \SI{3}{\arcminute} $ and $ w = 1\, \si{\mu\kelvin}\,\rm arcmin $.  The simulated spectra are then compared with theoretical ones using the a likelihood $\mathcal{L}$ \citep{Perotto:2006rj,Cabass:2015jwe}
\begin{equation}
	-2\ln\mathcal{L}_{\rm CMB} = \sum_{\ell} (2\ell + 1)f_{\rm sky}\left(\frac{D_\ell}{|C_\ell|} + \ln\frac{|C_\ell|}{|\hat{C_\ell}|} - 3 \right)~,
\end{equation} 
where $\hat{C} $ and $ C $ are the theoretical and simulated spectra with noise, respectively defined by 
\begin{align}
&|C_\ell
| = C_\ell^{TT}C_\ell^{EE}C_\ell^{BB} -
\left(C_\ell^{TE}\right)^2C_\ell^{BB}~;  \\
&|\hat{C}_\ell| = \hat{C}_\ell^{TT}\hat{C}_\ell^{EE}\hat{C}_\ell^{BB} -
\left(\hat{C}_\ell^{TE}\right)^2\hat{C}_\ell^{BB}~,
\end{align}
while $ D $ is
\begin{align}
	D_\ell  &=
	\hat{C}_\ell^{TT}C_\ell^{EE}C_\ell^{BB} +
	C_\ell^{TT}\hat{C}_\ell^{EE}C_\ell^{BB} +
	C_\ell^{TT}C_\ell^{EE}\hat{C}_\ell^{BB} \nonumber\\
	&- C_\ell^{TE}\left(C_\ell^{TE}\hat{C}_\ell^{BB} +
	2C_\ell^{TE}C_\ell^{BB} \right)~. \nonumber\\
\end{align}
The range of multipoles is assumed to be $ 5 \leq \ell \leq 3000 $ and the sky coverage of the $ 40\% $ ($f_{\rm sky} = 0.4$). We do not include CMB lensing derived from trispectrum data. 

For the future BAO dataset we instead consider the DESI experiment~\citep{DESI:2013agm} and we employ the volume average distance as a tracer for BAO~\footnote{It would also be possible to forecast BAO data considering $D_A/r_s$ and $H(z)$ as independent measurements, allowing for stronger constraints. However some small tension ($\sim 1 \sigma$) has been identified between the current constraints from $D_A/r_s$ and $H(z)$ \citep{Addison:2017fdm}. Therefore we follow the conservative approach of \citep{Allison:2015qca} and employ the volume average distance for the BAO forecasts.}
\begin{equation}
D_V(z)\equiv \left(\frac{(1+z)^2D_A(z)^2cz}{H(z)}\right)^\frac{1}{3}~,
\end{equation}
where $D_A$ is the angular diameter distance and $H(z)$ the Hubble parameter.
Assuming the very same fiducial model previously described, we compute the theoretical values of the ratio $D_V/r_s$ for several redshifts in the range $z=[0.15-1.85]$, where $r_s$ is sound horizon at the photon-baryon decoupling epoch. The uncertainties on $D_V/r_s$ are calculated propagating those for $D_A/r_s$ and $H(z)$ reported in \citep{DESI}. 
The simulated BAO data are compared to the theoretical $D_V/r_s$ values through a multivariate Gaussian likelihood 
\begin{equation}
    -2\ln \mathcal{L}_{\rm BAO} = \sum (\mathbf{\mu} - \hat{\mathbf{\mu}})C^{-1}(\mathbf{\mu} - \hat{\mathbf{\mu}})^T ~,
\end{equation}
where $\mu$ and $\hat{\mu}$ are the vectors containing the simulated and theoretical values of $D_V/r_s$ at different redshifts and $C$ their simulated covariance matrix.

\autoref{tab.forecast} summarizes the results obtained from our forecasting methods for future CMB and BAO experiments while \autoref{fig:forecast} shows the $68\%$ and $95\%$~CL contour plots for the different cosmological parameters. Using our forecasted data for future CMB-S4 observations, we derive the 68\% CL constraint $m_a=1.13\pm 0.11$ eV, showing that future CMB measurements will be able to reach a sensitivity on the axion mass $\sigma(m_a)\simeq 0.1$ eV; about one order of magnitude better than current Planck data. On the other hand, the limit on the total neutrino mass, $\sum m_{\nu}<0.178$ eV ($\sum m_{\nu}<0.220$ eV at 99\% CL), shows that when axions are included in the picture as additional thermal species, the possibility to detect the expected minimum neutrino mass of $\sum m_{\nu}\sim0.06$~eV with future cosmic probes is no longer possible and only upper bounds can be derived, as already pointed out in \cite{Giare:2021cqr}. Combining together the likelihood for CMB-S4 and DESI-like experiments we improve the constraint on the axion mass to $m_a=1.056^{+0.086}_{-0.075}$~eV, i.e. the expected sensitivity combining future CMB and BAO probes is $\sigma(m_a)\simeq 0.08$ eV. Similarly, the bound on the total neutrino mass is improved to $\sum m_{\nu}<0.118$ eV ($\sum m_{\nu}<0.144$ eV at 99\% CL), still below the accuracy needed for a combined measurement of the axion mass together with a detection of a neutrino mass. Notice however that we have explored a very conservative fiducial cosmology assuming the minimum neutrino mass allowed by oscillation experiments (i.e. $\sum m_\nu=0.06$~eV): larger neutrino masses could lead to smaller forecasted errors. Nevertheless the improvement in the measurement of the axion parameters in the context of future cosmological probes is highly remarkable, as can be noticed from a very quick comparison between \autoref{tab.forecast} and \autoref{tab.CMB.KSVZ} and \autoref{tab.CMB.DFSZ}.


\section{Conclusions}
\label{sec:conclusions}

The QCD axion is one of the strongest motivated particle for physics beyond the SM since it solves the strong CP problem and it provides a viable dark matter candidate. From a more general perspective, axion-like particles (ALPs) with properties similar to the QCD axion arise naturally in top-down motivated frameworks such as string compactification~\cite{Arvanitaki:2009fg}. Non-perturbative strong dynamics generates a non-vanishing axion mass, and when thermally produced axions become non-relativistic they leave cosmological signatures that are very similar to those originated by massive neutrinos. Therefore, constraints on axion properties must be derived in cosmological models including  also massive neutrino species.

Present constraints~\cite{Giare:2020vzo} on these hybrid hot dark matter scenarios rely upon approximate methods that could bias the constraints. On the one hand, they are derived by employing the instantaneous decoupling approximation and the resulting axion relic population is quite sensitive to the precise value of the decoupling temperature; this is due to the significant change in the effective number of the degrees of freedom of the thermal bath. The Boltzmann equation formalism avoids this problem because is not necessary to assume that thermal equilibrium is achieved in the early Universe, and it allows to track carefully the decoupling epoch. On the other hand, a delicate input in the Boltzmann equation is the axion production rate. Previous studies evaluated the cosmological axion mass bounds by considering the axion-pion and axion-gluon scattering separately. We employ here the results provided by Refs.~\cite{DEramo:2021psx,DEramo:2021lgb} in which the production rate is continuous across the QCD crossover. By means of the robust Boltzmann equation integration method, we revisit here the constraints in a hydrid mixed hot dark matter scenario, in which both neutrinos and QCD axions are present. 

We retrace the entire cosmological history of relic axions. First, we track their number density via a numerical solution of the Boltzmann equation and we evaluate their asymptotic comoving number density. This quantity gets frozen when axions stop experiencing any interactions with the thermal bath, and this happens always well above the MeV scale. Thus once we approach the first cosmological era that allows us to put meaningful constraints, namely the BBN epoch, the number of axions in the comoving volume is a fixed quantity that cannot change. We compare our predictions with observations probing the time of BBN (light element abundances), the recombination era (CMB temperature and polarization), and large scale structure (lensing and BAO).

For the axion mass range investigated here, axions are always relativistic at the time of BBN. They provide an additional contribution to dark radiation that we parameterize in terms of an effective number of additional neutrino species $\Delta N_{\rm eff}$. Analyzing light element abundances through BBN, we find the constraint $\Delta N_{\rm eff}<0.3$ at $95\%$~CL. For the KSVZ axion, we find the mass bound $m_a< 0.6$~eV (also at $95\%$~CL). If a prior on the baryon energy density from CMB is assumed, these bounds are further improved to $\Delta N_{\rm eff}<0.14$ and $m_a< 0.16$~eV, respectively. Similar results are found for the DFSZ axion (see \autoref{tab.BBN} for the complete set of BBN bounds).

When one considers later cosmological eras, axions may not be relativistic anymore because they lose their kinetic energy as the Universe expands. The recombination temperature is $0.3$~eV, and therefore axions heavier than approximately this scale (see \autoref{eq:TaLS}) are non-relativistic at the time of CMB formation. If this is the case, they do not contribute to dark radiation at recombination but rather provide a sub-dominant hot dark matter component. Furthermore, their non-vanishing mass is responsible for washing out cosmological structures at small scales through free-streaming. 

We employ cosmological observations from the CMB temperature, polarization and lensing from the Planck satellite combined with large scale structure data. In our analysis, we include the consequences of the axion presence at BBN. For the KSVZ axion, we find the bounds $\Delta N_{\rm eff}<0.23$, $m_a< 0.28$~eV and $\sum m_\nu < 0.16$~eV (all at $95\%$~CL). This corresponds to a factor of $\sim 5$ improvement in the axion mass with respect to the existing limits~\cite{Giare:2020vzo}. For the DFSZ axion the bounds are even more constraining, as a consequence of the larger production rate, and they explicitly read $\Delta N_{\rm eff}<0.24$, $m_a< 0.21$~eV and $\sum m_\nu < 0.16$~eV (all at $95\%$~CL). The full set of bounds from cosmology are summarized in \autoref{tab.CMB.KSVZ} and \autoref{tab.CMB.DFSZ} for the KSVZ and DFSZ axions, respectively.

Following the same procedure used for current observational data, we perform a forecasted analysis for upcoming cosmological and astrophysical surveys, targeting a CMB-S4-like observatory and a DESI-like BAO galaxy survey. For this analysis we chose a fiducial $\Lambda$CDM model with a non-vanishing axion and total neutrino mass, simulating constraints for the case of CMB-S4 alone and for CMB-S4+DESI. Our results show that future CMB observations will be able to reach a sensitivity on the value of the axion mass around an order of magnitude better than current constraints while the bounds on the total neutrino mass will be improved by a factor of $\sim 2$. When combined with a DESI-like experiment, we found a $\sim 20\%$ improvement in the constraints on the axion mass leading to $\sigma(m_a) = 0.08 $ eV (for a fiducial  $m_a = 1$~eV), while the total neutrino mass bound is improved by slightly more than $30\%$. 

Therefore, upcoming observations from future CMB and galaxy surveys could reach percent level constraints on the value of the axion mass. We refrain however that the inclusion of additional thermal species at early time significantly impacts the achievable bounds on the total neutrino masses preventing a direct measurements if $\sum m_\nu \lesssim 0.1$~eV.  

In concluding, while the future of the constraints on beyond the SM physics looks very bright from the point of view of next generation CMB and BAO experiments, extremely precise and carefully developed models of particle interactions in the early Universe are needed to keep theoretical modeling at pace with the improvements in accuracy that cosmological and astrophysical observations will perform in the current decade.  

\section*{CRediT authorship contribution statement}

\textbf{Francesco D’Eramo}: Conceptualization, Validation, Formal Analysis, Methodology, Writing - Original Draft, Writing - Review $\&$ Editing, Supervision.

\textbf{Eleonora Di Valentino}: Validation, Methodology,  Writing - Original Draft, Writing - Review $\&$ Editing, Supervision.

\textbf{William Giar\`e}: Conceptualization, Software,  Formal Analysis, Visualization, Validation, Methodology,  Writing - Original Draft, Writing - Review $\&$ Editing.

\textbf{Fazlollah Hajkarim}:  Conceptualization, Formal Analysis, Methodology,  Writing - Original Draft, Writing - Review $\&$ Editing.

\textbf{Alessandro Melchiorri }: Validation, Methodology,  Writing - Original Draft, Writing - Review $\&$ Editing, Supervision.

\textbf{Olga Mena}: Validation, Methodology,  Writing - Original Draft, Writing - Review $\&$ Editing, Supervision.

\textbf{Fabrizio Renzi}: Conceptualization, Software, Formal Analysis, Validation, Methodology,  Writing - Original Draft, Writing - Review $\&$ Editing.

\textbf{Seokhoon Yun}: Conceptualization, Formal Analysis, Methodology,  Writing - Original Draft, Writing - Review $\&$ Editing
\begin{acknowledgments}

The work of FD, FH and SY is supported by the research grants: ``The Dark Universe: A Synergic Multi-messenger Approach'' number 2017X7X85K under the program PRIN 2017 funded by the Ministero dell'Istruzione, Universit\`a e della Ricerca (MIUR); ``New Theoretical Tools for Axion Cosmology'' under the Supporting TAlent in ReSearch@University of Padova (STARS@UNIPD). EDV is supported by a Royal Society Dorothy Hodgkin Research Fellowship. FD, WG, FH, AM and SY are supported by Theoretical Astroparticle Physics (TAsP),``iniziativa specifica'' of Istituto Nazionale di Fisica Nucleare (INFN). The work of FD and OM is supported by the Spanish grants PID2020-113644GB-I00, PROMETEO/2019/083 and by the European ITN project HIDDeN (H2020-MSCA-ITN-2019/860881-HIDDeN. FR acknowledges support from the NWO and the Dutch Ministry of Education, Culture and Science (OCW) (through NWO VIDI Grant No.2019/ENW/00678104 and from the D-ITP consortium). FD acknowledges the hospitality while this work was in preparation of the Munich Institute for Astro- and Particle Physics (MIAPP) which is funded by the Deutsche Forschungsgemeinschaft (DFG, German Research Foundation) under Germany´s Excellence Strategy – EXC-2094 – 390783311
\end{acknowledgments}
\bibliography{main.bib}
\vfill
\end{document}